\documentclass[debug]{rmaa}
\usepackage{paralist}

\usepackage{psfrag,color}

\usepackage[latin1]{inputenc}

\usepackage{times}
\usepackage{amsmath}
\usepackage{mathptmx}
\usepackage{txfonts, ulem}
\usepackage{natbib}  
\usepackage{amssymb}

\usepackage{multirow}
\usepackage{adjustbox}
\usepackage{rotating}
\usepackage{txfonts}

\usepackage{booktabs}

\usepackage{lscape}

\newif\ifAMStwofonts
\AMStwofontstrue

\def\cm{\,{\rm cm}}

\def\ergscm2 {erg\,s$^{-1}$cm$^{-2}$}

\def\cm2 {cm$^{-2}$}

\def\aap {A\&A}
\def\apj {ApJ}
\def\apjs {ApJ Supplement Series}
\def\prd {Phys. Rev. D}
\def\prc {Phys. Rev. C}
\def\mnras {MNRAS}

\title{A resolution of the cosmic  Lithium problem} 
\author{Rachid Ouyed\altaffilmark{1}}
\altaffiltext{1}{Department of Physics \& Astronomy, University of Calgary, Canada}

\shortauthor{Rachid Ouyed}
\shorttitle{A plausible resolution of the cosmic  Lithium problem}
\fulladdresses{
\item Department of Physics \& Astronomy, University of Calgary, 
 2500 University Drive NW, Calgary, Alberta T2N 1N4, Canada (rouyed@ucalgary.ca).}
\listofauthors{Rachid Ouyed}
\indexauthor{Ouyed, R.}

\abstract{
In 1982,  Monique and Fran\c{c}ois Spite  discovered that the  $^7$Li abundance in  the atmosphere of old metal-poor  dwarf 
stars in the galactic halo was independent of metallicity and temperature. Since then, 
 $^7$Li  abundance in the Universe has become a subject of intrigue, because there is less of it in Population II dwarf  stars (by a factor  of 3) than standard big bang nucleosynthesis predicts. 
Here we show how quark-novae (QNe) occurring in the wake of  Pop. III stars,   can 
 elegantly produce an $A(Li)\sim 2.2$ Lithium plateau in Pop. II  (low-mass) stars formed 
  in the pristine cloud swept up by the mixed SN+QN ejecta. 
 We also find   an increase in the  scatter as well as an eventual  drop in $A(Li)$ below the Spite plateau values for  very low metallicity ([Fe/H]$<-3$) in excellent agreement with observations. 
     We propose a solution to the discrepancy between the Big Bang Nucleosynthesis $^7$Li abundance  and
    the Spite plateau and list some implications and predictions of our model.           
    }
    
    \resumen{---}
    
\addkeyword{Cosmology: Early Universe}
\addkeyword{Cosmology: Abundances}
\addkeyword{Stars: Population II}
\addkeyword{Stars: Population III}
\addkeyword{Supernovae: General }

\begin{document}
% Typeset article header
\maketitle

\section{Introduction} 

In 1982, Monique and Fran\c{c}ois Spite made a very important discovery. They found that the abundance of $^7$Li is essentially constant in warm metal poor stars in the galactic halo. Their derived abundance shows a plateau with a constant  $A(Li)$$\sim2.2$\footnote{We adopt the standard definition $[A/B] = \log{(N_A/N_{B})} - \log{(N_A/N_{B})_{\odot}}$
 and $A(Li)= \log{(N_{^{7}Li}/N_{H})}+12$. } for effective stellar temperatures of 5900 K $< T_{\rm eff}<$ 6300 K. This  $T_{\rm eff}$ range stands for a metallicity of -2.8 $<$ [Fe/H] $<$-2. The scatter in  $^7$Li abundance in these stars is really small, in the order of $\sim 0.05$ dex e.g. (e.g. Asplund et al. 2006).

	Although $^7$Li  is fragile to a relatively low temperature (e.g. $T \ge 2.5 \times 10^6$ K), warm, metal-poor stars (e.g. stars with temperatures of $T_{\rm eff}\ge$ 5900 K) have shallow convection zones, which prevents the destruction of $^7$Li. The star's atmosphere is not hot enough to destroy $^7$Li. If the convection zone is shallow, $^7$Li is preserved because it stays in the atmosphere and is not transported to the deeper, hot layers where it could get slowly destroyed\footnote{In standard stellar models, the expected depletion of $^7$Li for metal-poor turnoff stars is negligible, i.e., less than 0.02 dex (Deliyannis et al. 1990; Pinsonneault et al. 1992).}. The $^7$Li abundance in the atmosphere of these stars is expected to reflect the $^7$Li abundance of primitive Galactic matter. In these stars is ``tattooed"  the chemical composition of  the interstellar medium or cloud where they formed.

	 Originally, researchers proposed that this lithium plateau is the $^7$Li produced in the Big Bang, i.e. in primordial nucleosynthesis. However, theoretical and observational results challenge the idea (e.g. Ryan et al. 1999;  Cyburt et al. 2008). Through the Wilkinson Microwave Anisotropy Probe (WMAP) measurements of the cosmological constants, astronomers now know the abundances of BBN elements  ($^3$He, $^4$He, D, $^6$Li and $^7$Li ) produced by the Big Bang. All Big Bang Nucleosynthesis (BBN) elements were observed in the amounts predicted, except for   $^7$Li.  The corresponding primordial $^7$Li abundance derived is $A(Li)$ $ = 2.72^{+0.05}_{-0.06}$ (Cyburt et al. 2008; Coc \& Vangioni 2010), which is three times higher than the $^7$Li abundance observed in warm low-mass halo stars.

	 	Astronomers have discussed several ways to solve the discrepancy between the Spite plateau and the WMAP/BBN results. One idea involves the destruction of BBN $^7$Li through nuclear burning. Primordial, Population III stars deplete BBN lithium through nuclear burning, which leaves a lower lithium abundance that is eventually accreted into Spite plateau stars. 
		However, it appears that this idea cannot explain the entire BBN/Spite plateau gap (e.g. Sbordone et al. 2010). Furthermore, it appeals to Population III stars of 10-40M$_{\odot}$  which produce a larger yield of CNO elements than the observed abundances in metal-poor halo star (Prantzos 2007).

		Another resolution for the Spite/BBN discrepancy involves  non-standard BBN models  (e.g.
 Jedamzik et al. 2006; Coc et al. 2009; Iocco et al. 2009; Jedamzik \& Pospelov 2009;Kohri \& Santoso 2009). These models consider the decay of supersymmetric particles to lower the primordial prediction (e.g. Cyburt et al. 2010). For example, if the late decaying particle is a gravitino, then the Spite plateau may be reconciled with BBN values (e.g. Jedamzik et al. 2006). Yet, these non-standard models cannot explain the dependence of $A(Li)$ on metallicity below [Fe/H] $\sim-2.8$.

		A third solution involves the modification of reaction cross-sections for the production of $^7$Li in BBN (e.g. Chakraborty, Fields \& Olive 2011). However, ideas that involve the change of reaction -cross sections for $^7$Li were ruled out by nuclear and particle physics  (e.g. Coc et al. 2010). Furthermore, large errors in the cross section of the reaction leading to $^7$Li are unlikely (e.g. Cyburt et al. 2004; Angulo et al. 2005). 
		
		Finally, another idea involves the destruction of BBN lithium through its diffusion into a star's photosphere. Diffusion causes $^7$Li (and other elements) to sink below the photosphere where it is destroyed at  $T \ge 2\times 10^6$ K  by reactions with protons (e.g., Michaud et al. 1984; Salaris \& Weiss 2001). This depletion could in principle, attenuate the discrepancy between BBN and observed $^7$Li. The degree of this photospheric depletion can be parameterized by including mixing in the model (e.g. Lambert 2004). Modellers have applied mixing in diffusion through gravity waves (Talon et al. 2002; Charbonnel \& Talon 2005), rotation (e.g., Vauclair 1988; Chaboyer \& Demarque 1994; Pinsonneault et al.  2002; Talon \& Charbonnel 2004), and turbulence  (Richard et al. 2005; Piau 2008). Richard et al. (2005) managed to produce a flat $^7$Li abundance along the Spite plateau through a specific parameterization of turbulent mixing.

\subsection{The challenge}
\label{sec:challenge}

Any new model must face the following key issues  (see discussion in Spite et al. 2012 and references in Fields 2011): (i)  There is a very low scatter in the $^7$Li abundances among the Spite plateau stars.  $^7$Li depletion models face challenges when attempting to  reproduce this very low scatter ($\le 0.05$ dex); 
(ii)  The meltdown of the Spite plateau is established, but its cause is unclear (Sbordone et al. 2010).   Any successful model for the Spite plateau must be able to produce a constant lithium abundance among warm, metal-poor stars with [Fe/H] $>$ -2.8 that is metallicity independent and with low scatter. Furthermore, the model must be able to explain the metallicity dependent, high scatter lithium abundance in warm, metal poor stars with  [Fe/H] $<$ -2.8 as well. 

In this paper, we suggest an alternative solution. We argue that the Spite plateau might actually be produced by Quark-Novae (QN).  The QN is an evolutionary channel that is triggered after a type II supernova (SNII), where the neutron star (NS) left behind, if massive enough, experiences an explosive phase transition into a quark star  (Ouyed et al. 2002; Vogt et al. 2004; Ouyed et al. 2005; Ker\"anen et al. 2005; Niebergal et al. 2010). The explosion releases the neutron-rich outer layers of the parent NS, and these relativistic layers subsequently interact with the ejecta of the prior SNII explosion.  This neutron-rich ejecta  expands away from the quark star (the compact remnant) at relativistic speeds with an average Lorentz factor $\Gamma_{\rm QN}\simeq 10$; i.e. a typical neutron energy $E_{\rm QN} =10$ GeV (Ker\"anen et al. 2005; Ouyed et al. 2005).

	This interaction causes spallation reactions that produce new, daughter nuclei. In the case of the Spite plateau, we apply the QN model to Population III stars.  SNII explosions of Pop. III stars in the 20M$_{\odot}<$ M $<$ 40M$_{\odot}$ range should lead to massive NSs that evolve into QNe.  The QN's neutron-rich ejecta spall $^{56}$Ni and other SNII isotopes, which produces $^7$Li and other daughter nuclei. The fragmentation/destruction of $^{56}$Ni leads to iron-impoverishment in the cloud swept up by the combined SN+QN ejecta. These iron-poor clouds form into low metallicity Pop. II stars (Ouyed 2013). A key parameter is the time delay between the first SN explosion and the subsequent QN explosion. The metallicity is a function of $t_{\rm delay}$, where a low $t_{\rm delay}$ leads to a higher depletion of $^{56}$Ni and therefore a lower metallicity, and a high $t_{\rm delay}$ leads to a higher metallicity. We show that this primordial QN can lead to a production of a plateau of $A(Li)$$\sim 2.2$ and satisfy the criteria of low-scatter and appropriate melt-down. 

	The paper is organized as follows. In section 2, we briefly present the QN model. We explain the basic physics of the QN and discuss it in the context of Pop. III stars. Section 3 describes our model's parameters and how it relates to $^{56}$Ni destruction by neutron spallation and the resulting Fe impoverishment. In section 3, we also discuss how spallation results into sub-Fe nuclei and $\alpha$ elements, and compare this findings to observations. Section 4 describes how our QN model produces $^7$Li and a low scatter, plateau of $A(Li) \sim 2.2$. Section 5 discusses our results and predictions. We conclude in Section 6.
  
 \section{Primordial Quark-Novae}
 \label{sec:QN-model}
 
 The basic picture  of the QN is that a massive NS  (with $M_{\rm NS}> 1.6M_{\odot}$) converts 
explosively to a quark star.
Such an explosion can happen if the massive NS, in its spin-down evolution or via mass accretion (e.g.
from fall-back material), reaches the quark de-confinement density in its core  (Staff et al. 2006) and subsequently undergoes a phase transition to the conjectured more stable strange quark matter phase (Itoh 1970; Bodmer 1971; Witten 1984; see also Terazawa 1979). This results in a strange quark matter conversion front that propagates toward the surface of
the NS with a detonation occurring before the surface is reached
(Ker\"anen et al. 2005; Niebergal et al. 2010). The outcome of the QN explosion --  besides the formation of a quark star --  is the ejection of  the NS's outermost layers (a very neutron-rich
ejecta with an average mass $M_{\rm QN}\sim 10^{-3}M_{\odot}$) at  relativistic speeds (with a Lorentz factor averaging  $\Gamma_{\rm QN}\sim 10$; the evolution of this eject was analysed in details in Ouyed\&Leahy 2009).   The outer layers  are ejected  from an expanding thermal fireball (Vogt et al. 2004;
Ouyed et al. 2005)  which  allows for ejecta with kinetic energy, $E_{\rm QN}^{\rm KE} > 10^{52}$ erg.
  Hereafter we assume that a typical neutron in the QN ejecta will be streaming out with energy $E_{\rm QN}\simeq $ 10 GeV;
   not  to be confused with the total ejecta's kinetic energy $E_{\rm QN}^{\rm KE} $.
   This means that the $M_{\rm QN}\sim 10^{-3}M_{\odot}$ QN ejecta will have a typical total kinetic energy of 
   $E_{\rm QN}^{\rm KE} \sim 2\times 10^{52}$ erg.

  \subsection{Dual-shock Quark Novae (dsQNe)}
  
When the QN occurs  a few days to a few weeks 
   following the preceding Type II SN (SNII) explosion, a dual-shock QN (hereafter dsQNe) is created. The interaction between the relativistic, neutron-rich, QN ejecta 
 and the  SNII ejecta leads to spallation of the SN material  (Ouyed et al. 2011).
  This time delay, $t_{\rm delay}$, plays a crucial role in our model since it defines the
 density of the expanding SN ejecta (and thus spallation efficiency) when it is hit by the $E_{\rm QN}\sim 10$ GeV  QN neutrons (the primaries in our model; see \S \ref{sec:parameters} below).   If the time $t_{\rm delay}$  between SN and QN explosions  is too
 lengthy (of the order of months), the SN ejecta will have dissipated
such that the QN  erupts in isolation. However, when $t_{\rm delay}$ is on the order of days 
to weeks the neutron-rich QN ejecta interacts with the preceding SN ejecta leading to  spallation products that are
unique to dsQNe  (Ouyed et al. 2011; Ouyed 2013).

\subsection{The Pop. III Type-II SN (PopIII-SNII) progenitor}
\label{sec:progenitor}

Studies of star formation in the early Universe suggest that
the first stars (Pop. III) were  more massive than present day
objects, with a typical mass of $\sim 100M_{\odot}$  for zero metallicity ($Z$) objects
(Abel et al. 2002; Nakamura \& Umemura
2002;  Bromm\&Larson 2004;  Loeb et al. 2008). The progenitors of QNe are  stars with masses in the   $20M_{\odot} < M_{\rm prog.} < 40M_{\odot}$ range (e.g. Ouyed et al. 2009a;
Ouyed et al. 2009b;  Leahy\&Ouyed 2009; Ouyed\&Leahy 2013; Ouyed 2013). 
Thus, primordial QNe are expected to occur in the wake of these Pop. III stars from progenitors
with mass range  at the lower end of the mass distribution.
 These are expected to explode as TypeII SNe and would most likely form NSs
  massive enough to result in  QNe once quark de-confinement density is reached in their cores (Staff et al. 2006).
  Hereafter we refer to them as PopIII-SNII progenitors.

\section{Our model}

  In this section,  we summarize and discuss our model's parameters and its general features and 
 present some results. As in our previous studies,    in all of the calculations 
 presented here, we divide the target layer into $N_{\rm mfp, A_T}$ sublayers
 of radial thickness equivalent to a spallation mean-free-path (mfp), $\lambda_{\rm sp., A_T}$. 
 In each mfp of interaction, a given nucleus with atomic number  $A_{\rm T}$ 
  will be hit multiple times by the neutrons. To produce a realistic distribution of
   product nuclei from this sub-layer, we draw  the number of hits from a 
    Poisson distribution.     The interested reader is referred to 
 Ouyed et al. (2011; see eqs 3, 4 and 5 in that paper) and Ouyed (2013) for more details
 on the  spallation process, numerical procedure, and relevant references.
       
\subsection{Model Parameters}
\label{sec:parameters}

 We start by summarizing the  key parameters in our model:

 \begin{itemize}
 
 \item {\bf The PopIII-SNII ejecta's composition}:   Studies of the chemical composition of PopIII-SNII ejecta in the $20M_{\odot} < M_{\rm prog.} <  40M_{\odot}$ mass range can be found
 in the literature for mixed and unmixed SN ejecta
 (Umeda\&Nomoto 2002; Nozawa et al. 2003; Cherchneff\&Dwek 2009; Heger\& Woosley 2010).
 Among the main products of these PopIII-SNII supernovae as listed in these studies are 
   $^{56}$Ni,$^{32}$S, $^{28}$Si,$^{24}$Mg,$^{20}$Ne,$^{16}$O and, $^{12}$C.  
 Other elements (including $^{14}$N; see \S \ref{sec:nitrogen} below)
are present at much smaller levels so we ignore them here.  
We also ignore elements heavier than $^{56}$Ni in the present study.   
While some  differences exist
 in mass yields arrived by  different studies/groups, and between models with mixing and non-mixing, we find that a typical $20M_{\odot}$ PopIII-SN ejecta (with  $10^{51}$ erg in energy) can be
 represented with an averaged mass yields  of $M_{\rm Ni, SN} \sim 0.1M_{\odot}$, $M_{\rm S, SN} \sim 0.05M_{\odot}$, 
 $M_{\rm Si, SN}\sim 0.05M_{\odot}$,  $M_{\rm Mg, SN} \sim 0.05M_{\odot}$, $M_{\rm Ne, SN}\sim 0.05M_{\odot}, 
M_{\rm O, SN}\sim 1.5M_{\odot}, M_{\rm C, SN}\sim 0.15M_{\odot}$.  We adopt
these values as initial, fiducial, values for the target material; i.e. the PopIII-SN ejecta in our model.

Although some mixing is likely to occur during the PopIII-SNII explosions, here
  we  assume an onion-like profile of the expanding  SN ejecta (i.e., no mixing) with the innermost ejecta, viz., $^{56}$Ni nuclei (mass number $A_{\rm T}$=$56$)  constituting the target $A_{\rm T}$ at a distance $R_{\rm  A_T}(t_{\rm delay})$=$ v_{\rm sn}\,t_{\rm delay}$ from the center of the explosion. Furthermore,  we point out that  we assume $v_{\rm sn}$ to be constant since the QN occurs while the SN ejecta is still in the Sedov phase (e.g., McCray 1985).   Carbon ($^{12}$C) makes up the last layer in this onion-like geometry overlaying the Oxygen layer.
 The $^4$He and H layers overlaying the carbon layer can be ignored since in most cases
considered here we find that  the spallation mean-free-path of the (primary, secondary and subsequent generations of) neutrons  always exceeds the  (H and $^4$He) layers'
thickness.

 \item  {\bf The time delay ($t_{\rm delay}$)}: The number density in the target layer  (made of element with atomic number $A_{\rm T}$)  for the case of a uniform density is 
    $n_{\rm A_T}= M_{\rm A_T, SN}/(4\pi R_{\rm A_T}^2 \Delta R_{\rm A_T} A_{\rm T} m_{\rm H}$) where 
  $M_{\rm A_T, SN}$ is the mass of the layer (the target's initial mass in the
    PopIII-SN ejecta prior to impact by the QN neutrons), $\Delta R_{\rm A_T}$ the layer's thickness
   and $R_{\rm A_T}= v_{\rm sn} t_{\rm delay}$ the radius of the layer from the center of
    the SN explosion; $m_{\rm H}$ is the hydrogen mass.  For a given PopIII-SNII explosion with known  $M_{\rm A_T, SN}$ (e.g.
     for a fixed amount of $^{56}$Ni processed during the explosion) and given $v_{\rm sn}$,  which we assume to remain 
     constant early in the expansion of the SN ejecta, $t_{\rm delay}$ becomes the key parameter
      that defines the target density. In this prescription,  the longer the time delay, $t_{\rm delay}$, the lower
     the target density when it is hit by the relativistic QN neutrons.  
    
     The neutron  spallation mfp on a target $A_{\rm T}$ is $\lambda_{\rm sp., A_T}= 1/(n_{\rm A_T} \sigma_{\rm sp., A_T})$
     where the spallation cross section in milli-barns is given by  $\sigma_{\rm sp., A_T} = 45 A_{\rm A_T}^{0.7}$ mb (see  Ouyed et al. 2011).
     The average number of collisions (i.e. the number of spallation mfps) an incoming neutron experiences in the target
     layer $A_{\rm T}$ is then 
\begin{equation}
 N_{\rm mfp, A_T}\approx\frac{\Delta R_{\rm A_T}}{\lambda_{\rm sp., A_T}}\simeq 0.69 \frac{M_{\rm A_T,SN}/0.1M_{\odot}}{{\left(A_{T, 56}\right)}^{0.3}(v_{\rm sn, 5000}\, t_{\rm delay, 10})^2} \ ,
\label{eq:Nsp}
\end{equation}
where  $A_{\rm T,56}$ is the target atomic mass in units of $56$ and $0.1M_{\odot}$ representative of the mass in the $^{56}$Ni  layer. In
    the equation above, $v_{\rm sn}$ is given in units of 5000 km s$^{-1}$ and is kept constant in all
of the simulations shown in this paper. The time delay, $t_{\rm delay}$, is in units of 10 days. For our fiducial values in mass
 of the target material and  for $t_{\rm delay} = 10$ days we get
$N_{\rm mfp, Ni}\sim 0.69, N_{\rm mfp, S}\sim 0.41, N_{\rm mfp, Si}\sim 0.42, N_{\rm mfp, Mg}\sim 0.44,  N_{\rm mfp, Ne}\sim 0.47, N_{\rm mfp, O}\sim 15.10, N_{\rm mfp, C}\sim 1.64$ 
for  spallation on $^{56}$Ni,   $^{32}$S,   $^{28}$Si,   $^{24}$Mg,   $^{20}$Ne,   $^{16}$O,
$^{12}$C, respectively.

     In a regime where  $\lambda_{\rm sp., A_T} > \Delta R_{\rm A_T}$ (i.e. $N_{\rm mfp, A_T} < 1$), there will be minimal
     or even no spallation induced by the QN neutrons on the target $A_{\rm T}$ in which case most of the neutrons proceed
     onto the next, overlaying, layer. Thus, for a given 
     $M_{\rm A_T, SN}$ and $v_{\rm sn}$  this defines a critical time delay  above which
     no spallation occurs in layer of element $A_{\rm T}$.
      This allows us to define  four regimes in our model:
      
       (i)  $t_{\rm delay} > t_{\rm outer}$ with minimal or no spallation occurring in the inner layers.    It means that 
  for $t_{\rm delay} > t_{\rm outer}\simeq 8.3\ {\rm days}\times (M_{\rm Ni, SN}/0.1 M_{\odot})^{1/2}/(A_{\rm T, 56}^{0.15} v_{\rm sn, 5000})$ days (see eq. 2 in Ouyed 2013),  the QN neutrons barely interact with the inner layers and proceed
directly to the outer (O and C layers);  

(ii) $t_{\rm delay} < t_{\rm inner}$ with spallation occurring mainly in the inner  layers, beneath the oxygen layer,  with  $t_{\rm inner}\simeq 6.3\ {\rm days}\times (M_{\rm Ni, SN}/0.1 M_{\odot})^{1/2}/(A_{\rm T, 56}^{0.15} v_{\rm sn, 5000})$ (see Eq. 6 in Ouyed 2013)\footnote{In Ouyed (2013), $t_{\rm outer}$ and $t_{\rm inner}$
were defined as $t_{\rm Ni}$ and $t_{\rm O}$, respectively.}.
  This regime corresponds to  the situation
 where the  spallated (i.e. subsequent generations of) neutrons exiting the inner layers, do not have enough energy to spallate
 the outer CO-rich layers;
 
   (iii) $t_{\rm inner} < t_{\rm delay} < t_{\rm outer}$  with spallation occurring both in the inner and outer layers.
 In this regime, the neutrons spallated
  in the inner layers end up with  energies above the critical values for spallation in the $^{16}$O (and in certain
   cases in the $^{12}$C) to ensue;

 (iv)  $t_{\rm delay} > t_{\rm no-spall}$ which effectively defines a regime  
 of a QN occurring in isolation since  no interaction occurs between the SN and QN ejecta.
 For our fiducial values, $t_{\rm no-spall} \sim $23-24 days. For $t_{\rm outer} < t_{\rm delay} < t_{\rm no-spall}$,
  spallation would occur mainly in the oxygen and carbon layers.

   \item  {\bf The energy   per nucleon, $E_{\rm QN}$,  in the neutron-rich QN ejecta}: 
     It defines and sets the total multiplicity, $\zeta$, of the spallated (i.e. subsequent generations of) neutrons and protons 
      and their  energies. The  average neutrons$+$proton multiplicity $\zeta_{\rm av.}$ on target $A_{\rm T}$
       is given in eq. (2) in Ouyed (2013) as
      \begin{equation}
      \zeta_{\rm av.} \simeq 7 A_{\rm A_T, 56} \times (1+ 0.38 \ln{E_{\rm n}})\ ,
      \label{eq:zeta}
      \end{equation}
      where $E_{\rm n}$ is the neutron energy in GeV which leads to a subsequent generation
      of spallated neutrons and protons with energy $\sim E_{\rm n}/\zeta_{\rm av.} $. The resulting spallation product's atomic weight will peak at $A_{\rm T}-\zeta_{\rm av.}$.
      The $ \zeta_{\rm av.} > 1$ condition gives  the minimum neutron   energy  necessary for spallation to occur in the
        Ni, S, Si, Mg, Ne, O, C layer  to be  $E_{\rm sp., A_T}\sim 0.105, 0.139, 0.153, 0.173, 0.206, 0.268, 0.416$ GeV,  respectively.

        Table \ref{tab:ni-spallation} shows results from simulations involving the interaction of $E_{\rm QN}=5$ GeV and  $E_{\rm QN}=10$ GeV neutrons
        with the PopIII-SNII ejecta when  $t_{\rm delay}=t_{\rm outer} \simeq 8.3$ days.  For our fiducial values of the targets'
         initial masses this time delay implies $N_{\rm mfp, Ni}\sim N_{\rm mfp,  S}\sim N_{\rm mfp,  Si}\sim N_{\rm mfp,  Mg}\sim N_{\rm mfp,  Ne}\sim 1$
         which means that  on average a QN neutron will interact once per layer.    
           An $E_0=E_{\rm QN}=$ 10 GeV  QN (primary) neutron would lead to a multiplicity of $\zeta_{\rm av.} \sim 13$ in the $^{56}$Ni layer.
          This  means a   spallated (secondary generation of) neutrons  and protons  exiting the layer with energy $E_1 =E_{\rm 0} /\zeta_{\rm av.}  \sim 0.77$ GeV.  These secondary nucleons induce   an average multiplicity $\zeta_{\rm av.} \sim 3.6$ in the $^{32}$S layer which yields 
       spallated (tertiary generation of) neutrons  and protons exiting the layer with energy $\sim 0.21$ GeV.
          As can be seen from Table \ref{tab:ni-spallation},  spallation   ceases at the Si layer since within  the first three layers the neutrons energy  has cascaded down to  $\sim 0.15$ GeV,  below
          the critical value for Mg-spallation.   The main (i.e. peak) spallation products are shown as well as the net multiplicity $\zeta_{\rm net}$ which yields a 
         total mass of spallated neutrons and protons which amounts to $M_{\rm n+p}= \zeta_{\rm net}M_{\rm QN}\sim$ 0.05-0.1M$_{\odot}$.        This table shows the pathway of a  typical neutron with the expected average multiplicity given by
           eq. (\ref{eq:zeta}). However,  because the $\zeta$ distribution is not uniform,  some neutrons  induce above average multiplicity ($\zeta > \zeta_{\rm av.}$) and stop within the $^{56}$Ni layer 
           while those inducing below average multiplicity ($\zeta < \zeta_{\rm av.}$) make it past the inner layers and induce spallation in the oxygen layer (see discussion below).

        Another example is shown in Table \ref{tab:o-spallation} which shows results from the interaction of $E_{\rm QN}=5$ GeV and  $E_{\rm QN}=10$ GeV neutrons
        with the PopIII-SNII ejecta when  $t_{\rm delay} > t_{\rm outer}$ (here $t_{\rm delay}=12$ days). For our fiducial values, $N_{\rm \lambda_{sp.}, O}\sim 10$ and $E_{\rm sp., O}\sim 0.27$ GeV.
        As can be seen from Table \ref{tab:o-spallation}, neutrons interact with five sub-layers (each representing a spallation mfp) before their energy cascade down below 0.268 GeV.  Thus half the 
        amount of oxygen in the PopIII-SNII will be destroyed forming sub-O products.         
           In each sub-layer denoted by index $i$ (i.e. for each mfp, or $\lambda_{\rm sp., O}$),  the atomic weight of the spallation products  is close to 
          a normal distribution peaking at atomic weight $A_{\rm peak, i}\sim A_{\rm T}- \zeta_{\rm av., i}$.
           The final abundances of the spallation products is a combination of these distributions (see Figure 1 in Ouyed et al. 2011).  As can be seen in Table 2 the total spallation neutrons and protons amounts to $M_{\rm n+p}= \zeta_{\rm net}M_{\rm QN}\sim 0.02$-0.04 M$_{\odot}$.  
  This Table further illustrates the important point that most of the
  spallated  neutrons exiting the oxygen layer will not be able to induce spallation in the overlaying C layer since their energy
 is below the threshold for C-spallation ($\sim 0.42$ GeV).  See however discussion below.

          We end this subsection by  mentioning some  important points:\\

        (i)  The true $\zeta$ distribution is not uniform and is rather close to a normal distribution peaking at $\zeta_{\rm av.}$.
         This means that  the spallated 
  neutrons would exit a sub-layer $i$ of  the target  material ($A_{\rm T}$) with a non-uniform energy distribution peaking at $\sim E_{\rm i}/\zeta_{\rm av., i}$.      For the case shown in Table  \ref{tab:o-spallation},   most of the neutrons spallated in the 
   oxygen layer will not be able to induce spallation in the overlaying C layer -- since their energy
 is below the threshold for C-spallation $E_{\rm sp., C} \sim 0.42$ GeV. However,  neutrons 
 leading to   a multiplicity $\zeta < \zeta_{\rm av.}$  do induce spallation after exiting the oxygen layer.  As we show later (see \S \ref{sec:c-plateau}), 
 this has important consequences to production of $^7$Li from
  spallation in the C-layer;
        
        (ii) The logarithmic energy dependence of $\zeta_{\rm av.}$ means that there is  little difference between
             the multiplicity distribution induced by the $E_{\rm QN}=10$ GeV and the  $E_{\rm QN}=5$ GeV neutrons. However,
              for the lower energy  primaries  with   $E_0=E_{\rm QN}$ (i.e. QN neutrons), the resulting average energy per spallated nucleon  is lower. In the case of spallation in the Oxygen layer for example (as illustrated
              in Table  \ref{tab:o-spallation}), the $E_{\rm QN}=5$ GeV simulations shows  
   fewer neutrons with  energy above $E_{\rm sp., C}\sim 0.42$ GeV making it past the oxygen layer;
  i.e. spallation in the overlaying C-layer is reduced  as compared to the $E_{\rm QN}=10$ GeV simulations (discussed
  further in \S \ref{sec:c-plateau});

          (iii) Even if the  spallation condition, $E_{\rm n} > E_{\rm sp., A_T}$,  is satisfied in any layer of
      target material $A_{\rm T}$ (a necessary
        condition), it is not  sufficient when $\lambda_{\rm sp., A_T} > \Delta R_{\rm A_T}$; i.e.  when the spallation
        mfp exceeds the thickness of the target layer (see discussion  following eq. \ref{eq:Nsp}). 
        Thus the $E_{\rm n} > E_{\rm sp., A_T}$ and  the $ \Delta R_{\rm A_T} > \lambda_{\rm sp., A_T}$ 
         must be satisfied simultaneously.

\end{itemize}

\subsection{$^{56}$Ni destruction and Iron-impoverishment in our model}

\subsubsection{The $t_{\rm delay}$-[Fe/H] connection}
\label{sec:tdelay-FeH}

Following spallation in the 
 $^{56}$Ni layer, the resulting iron abundance ([Fe/H])
 in the primordial cloud swept by the combined SN$+$QN ejecta is (eq. 5 in Ouyed 2013): 
\begin{equation}
\left[ \frac{Fe}{H}\right] =  \log{\eta_{56}^{56}} + \log{\frac{M_{\rm Ni, SN}/0.1M_{\odot}}{M_{\rm sw}/ 10^5 M_{\odot}}} - 3.12\ ,
 \label{eq:FeH}
\end{equation}
where $\eta_{\rm A}^{A_{\rm T}} = M_{\rm A}/M_{\rm A_{\rm T}, SN}$ is the normalized mass yields of product/element $A$
following spallation on target $A_{\rm T}$ (originally present in the SN ejecta; here 56 refers to $^{56}$Ni). Here $M_{\rm sw}$ is the mass
swept up by the mixed SN$+$QN ejecta in the primordial pristine cloud surrounding the site of the dsQN explosion; $M_{\rm sw}$
is in units of $10^5M_{\odot}$ (e.g. Shigeyama\&Tsujimoto 1998; Machida et al. 2005).  

In our model,  a dsQN with a given $t_{\rm delay}$ translates to a corresponding  [Fe/H] abundance in the swept up cloud;
 this is set by the parameter  $\eta_{56}^{56}$ which is the  level of 
 depletion of the original PopIII-SNII $^{56}$Ni  following spallation.
 Figure 1 in Ouyed (2013) shows the heavy  depletion of the original PopIII-SNII $^{56}$Ni by spallation
  obtained for very short time delays ($t_{\rm delay}<$ 2 days)
  where $\eta_{56}^{56} <  10^{-4}$ (i.e.  leading to [Fe/H]$< -7$) . 
 Hyper metal-poor stars (HMPs with [Fe/H]$<-5$ in the terminology of Beers\&Christlieb 2005)  
   would form from clouds swept up by $t_{\rm delay} < 3$ days dsQNe in our model (see \S \ref{sec:discussion}). 

In dsQNe with long enough delays (i.e. $t_{\rm delay} > t_{\rm outer}$),   there is minimal spallation
in the $^{56}$Ni layer (i.e. $\eta_{56}^{56}\simeq 1$) and in the innermost layers ($^{32}$S, $^{28}$S, $^{24}$Mg) layers. For such  dsQNe,  high values of [Fe/H]  are obtained by increasing $M_{\rm Ni, SN}$
     or by decreasing the amount of swept up material, $M_{\rm sw}$ (see eq. \ref{eq:FeH} above). 
For example, a dsQN with $t_{\rm delay} > 10$ days, $M_{\rm Ni, SN}= 0.5M_{\odot}$ and $M_{\rm sw}=10^{4.5} M_{\odot}$ gives [Fe/H]$\sim - 1.9$.

    \subsubsection{Sub-Fe elements}
    \label{sec:subFe}

\begin{figure*}
\begin{center}
\begin{tabular}{cc}
\includegraphics[height=0.3\textwidth,width=0.5\textwidth,]{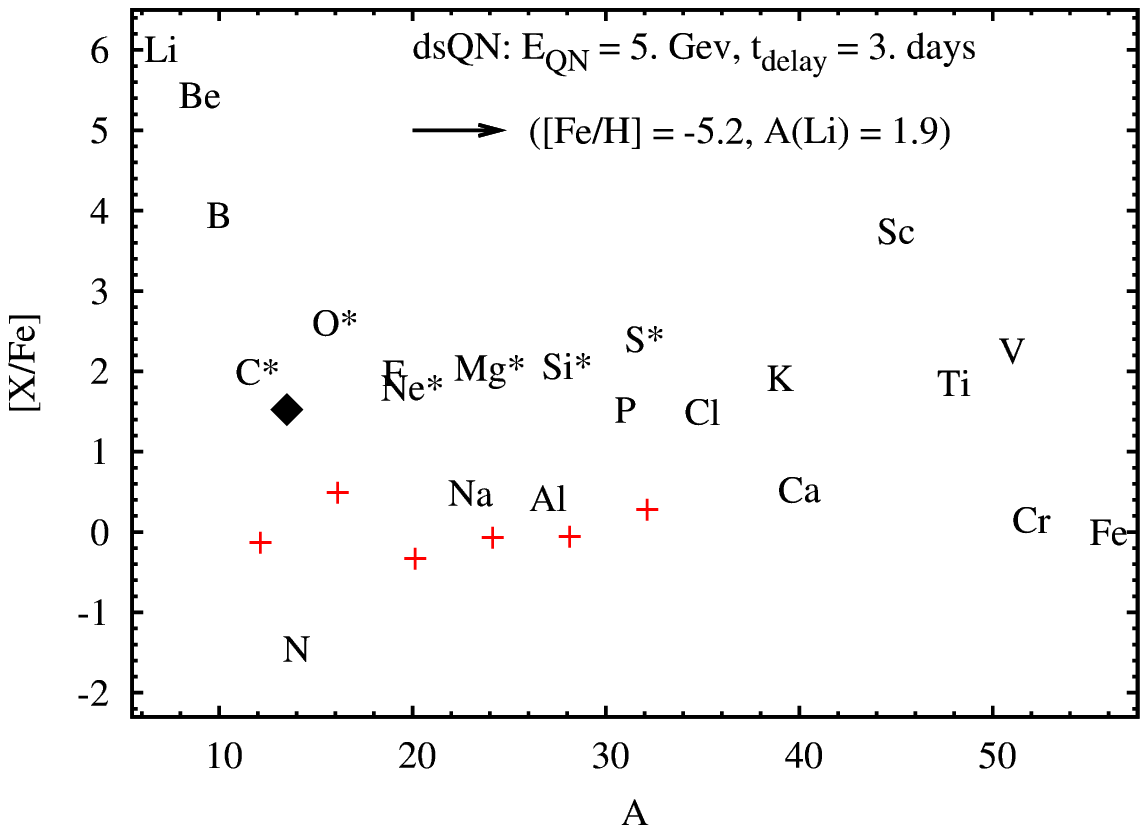} &  \includegraphics[height=0.3\textwidth,width=0.5\textwidth,]{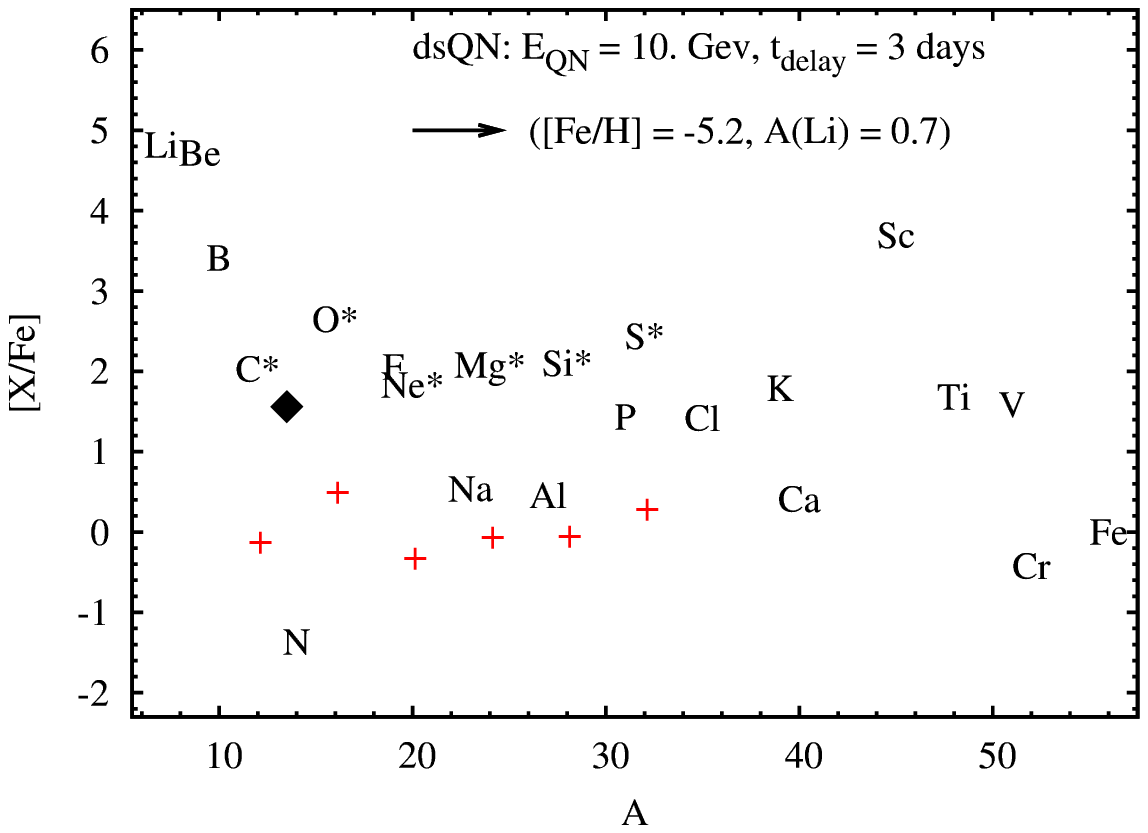} \\
\includegraphics[height=0.3\textwidth,width=0.5\textwidth,]{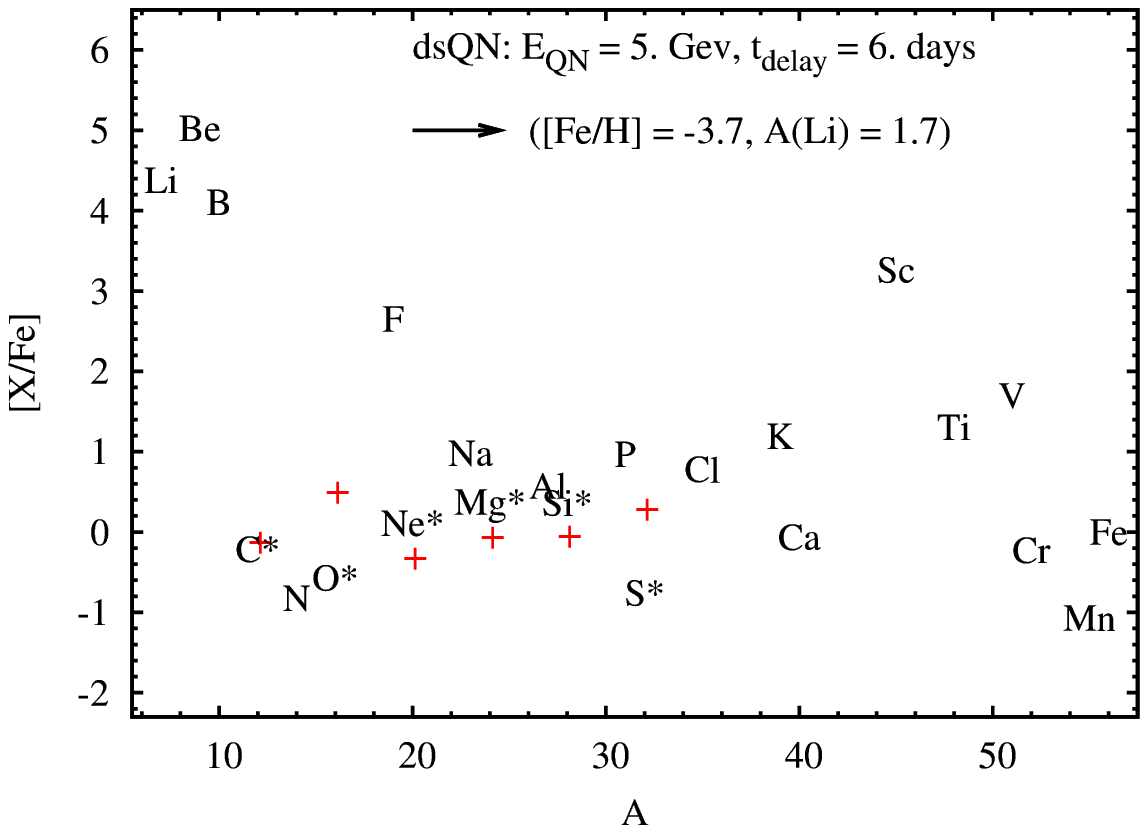} &  \includegraphics[height=0.3\textwidth,width=0.5\textwidth,]{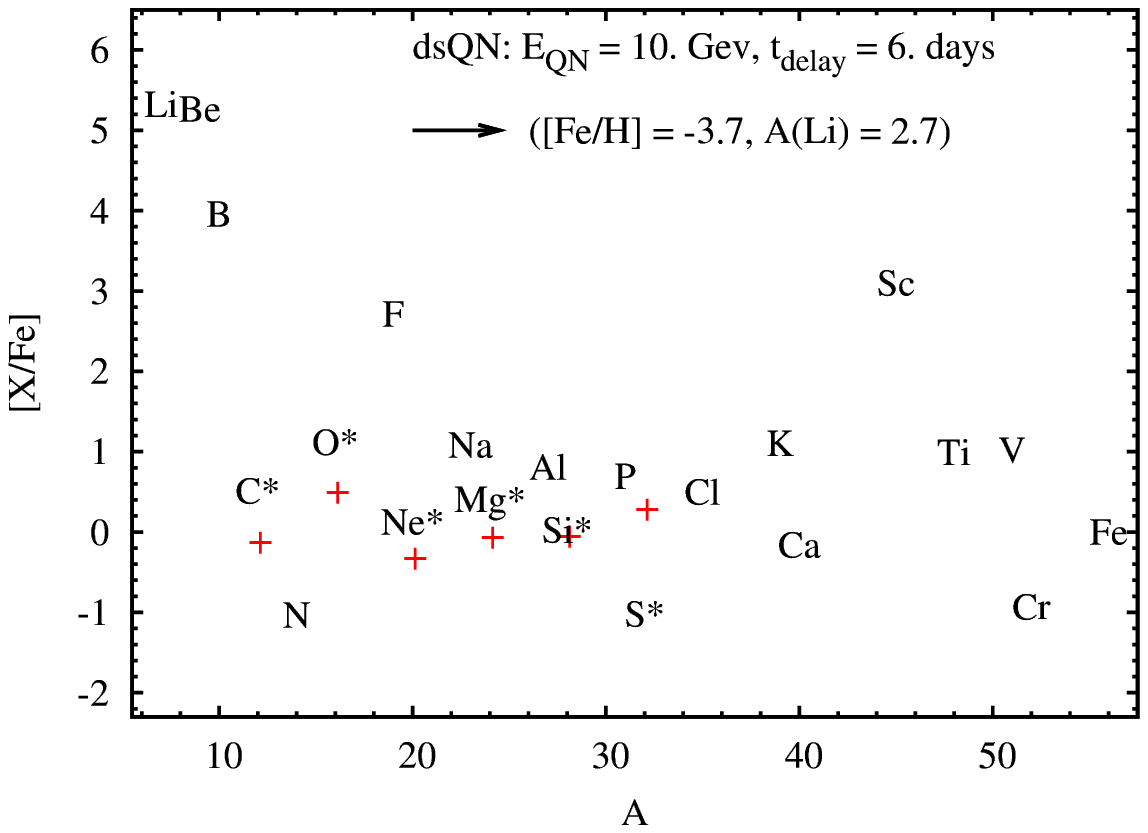} \\
\includegraphics[height=0.3\textwidth,width=0.5\textwidth,]{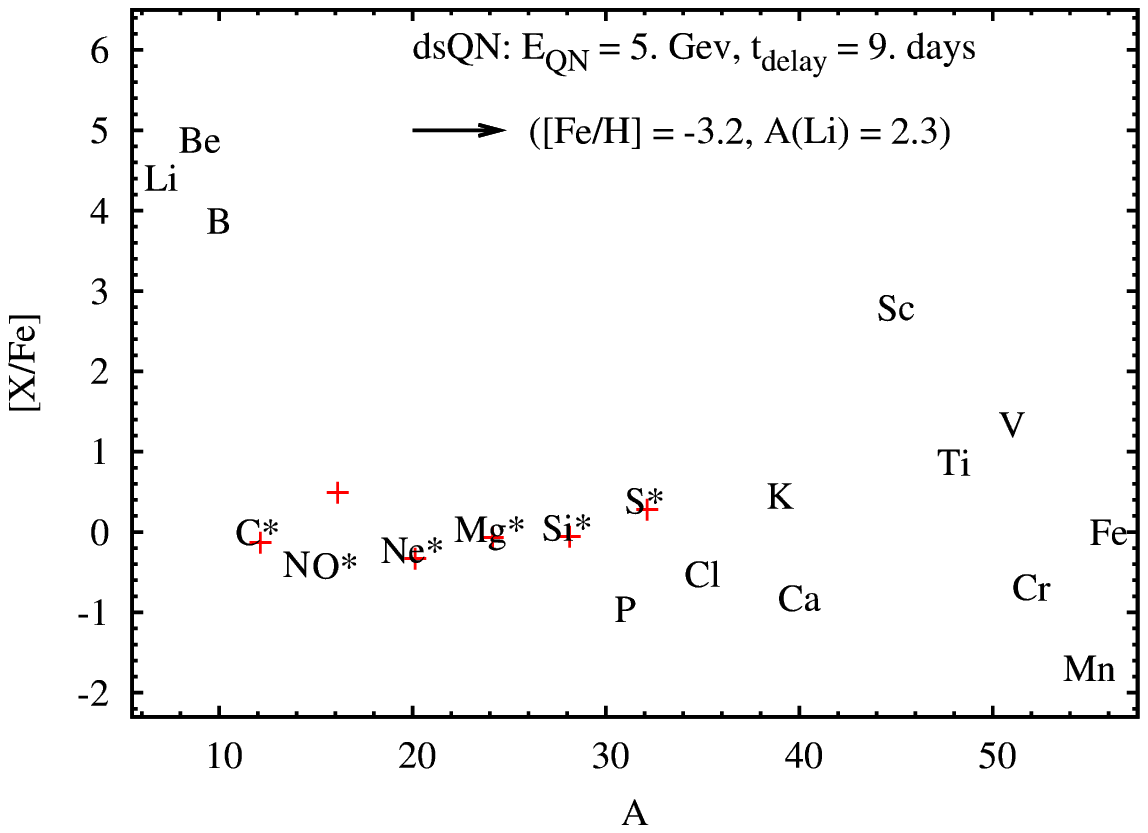} &  \includegraphics[height=0.3\textwidth,width=0.5\textwidth,]{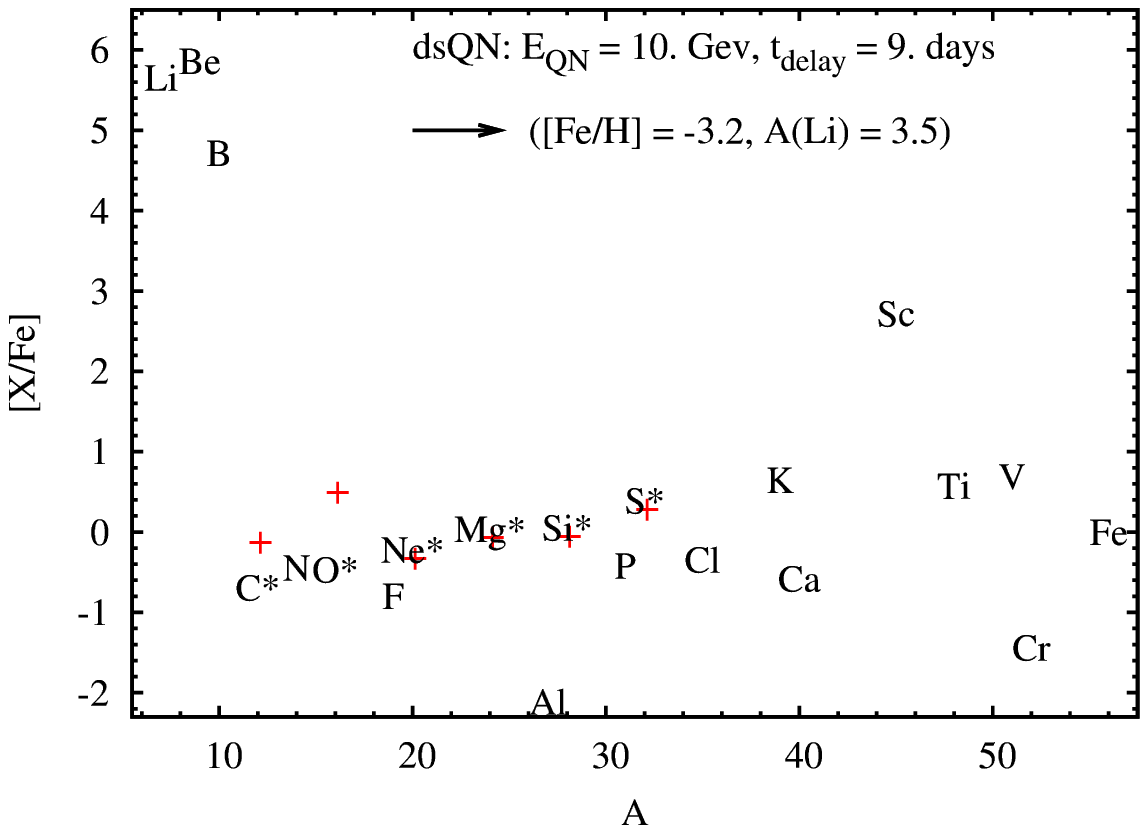} \\
\includegraphics[height=0.3\textwidth,width=0.5\textwidth,]{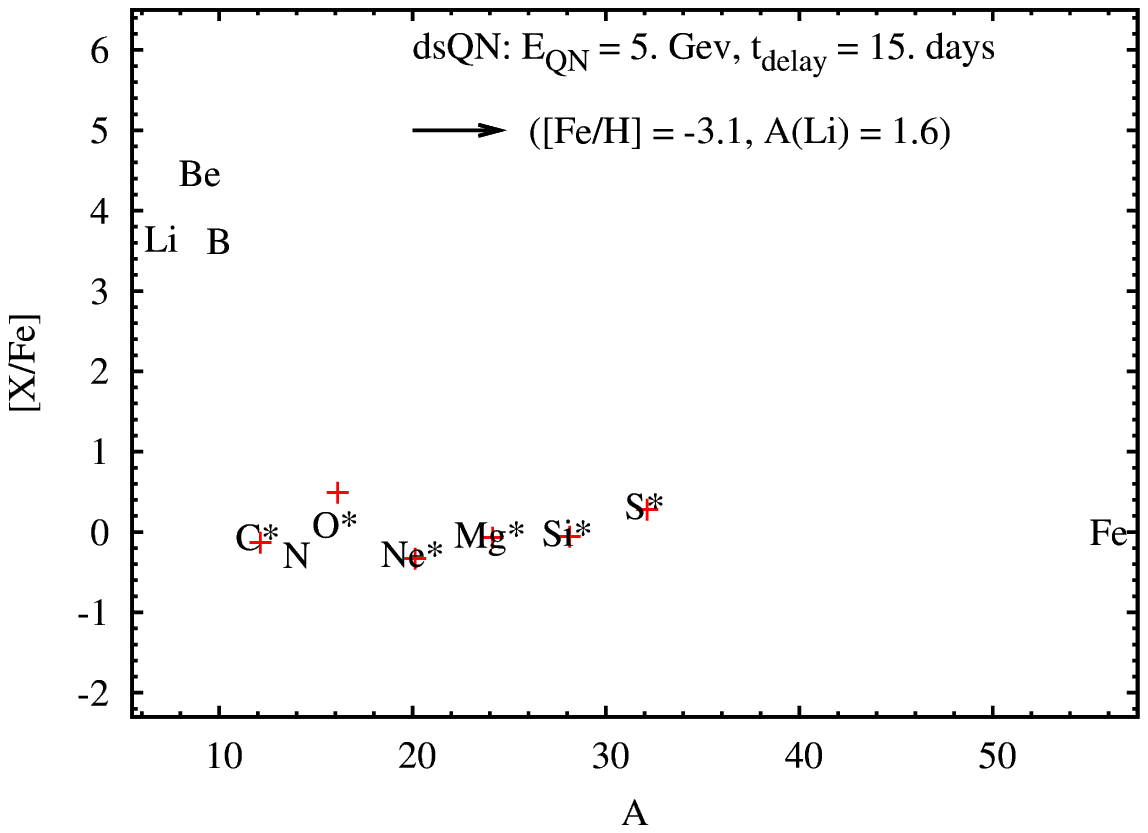} &  \includegraphics[height=0.3\textwidth,width=0.5\textwidth,]{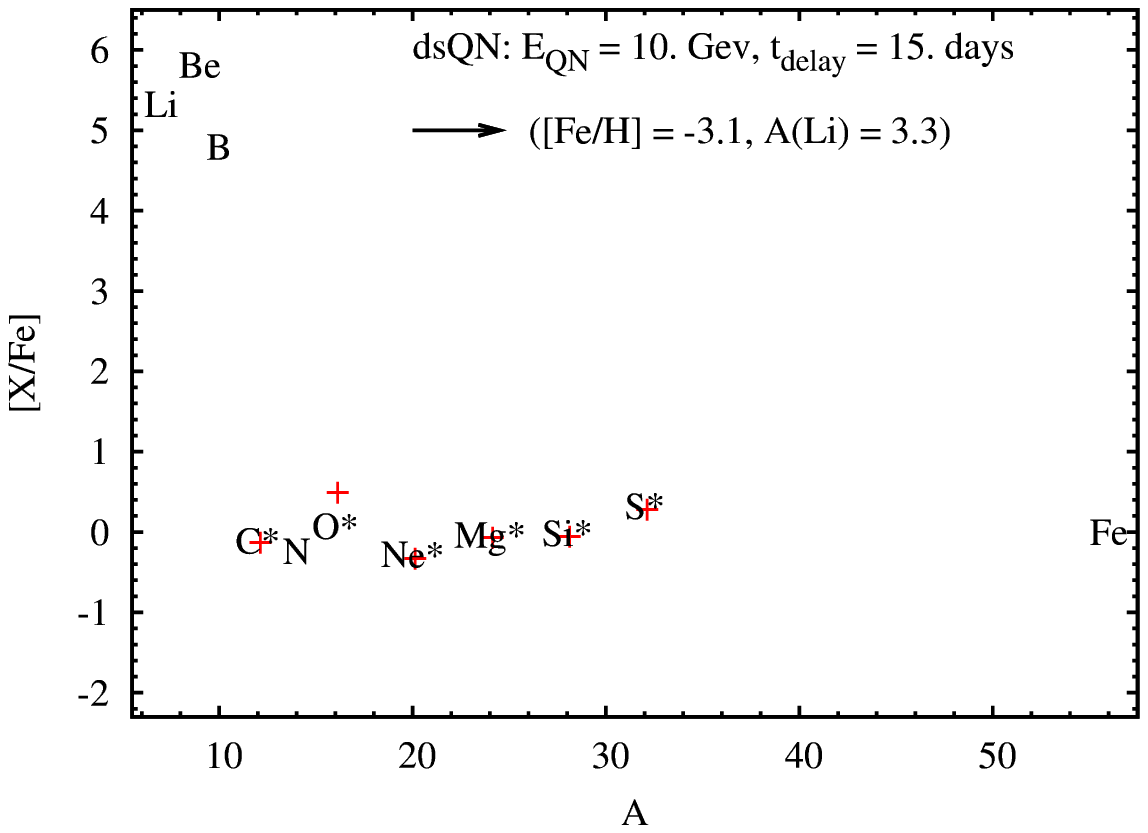} \\
\end{tabular}
\caption{Relative abundances [X/Fe]  for spallation (sub-Fe) products versus mass number $A$ ($ 7 \le A < 56$;
only stable isotopes are shown here and in other figures in this paper) for a dsQN with 
 5 GeV (left panels) and 10 GeV (right panels) QN neutrons.  Letters show abundances from our simulations with
  the CN enhanced nitrogen (for short $t_{\rm delay}$) shown as a filled diamond. The panels 
 from top to bottom are for  $t_{\rm delay} ({\rm days})=3, 6, 9, 15$, respectively.
 Also shown  are the resulting [Fe/H]  and  $^7$Li
 abundance, $A(Li)$. The plus signs represents the initial relative abundance of the
 original PopIII-SNII elements; these elements  are identified  by an asterix (next to their names) as they evolve with $t_{\rm delay}$. 
 Their  initial abundances in mass  are  $M_{\rm Ni, SN} =0.1M_{\odot}$, $M_{\rm S, SN}=M_{\rm Si, SN}=M_{\rm Mg, SN}=M_{\rm Ne, SN}= 0.05M_{\odot}$, $M_{\rm O, SN}=1.5M_{\odot}$ and $M_{\rm C, SN}= 0.15M_{\odot}$.   These initial abundances are the same in all of the figures  in this paper unless stated otherwise.   Similarly, unless stated otherwise, the mass of the cloud swept by a dsQN
  is set to $M_{\rm sw}=10^{5}M_{\odot}$ in all of our simulations.
}
\label{fig:X-over-Fe}
\end{center}
\end{figure*}

  Shown in Figure \ref{fig:X-over-Fe}
  in this paper are  abundances of sub-Fe spallation products resulting from dsQNe with different $t_{\rm delay}$;
  shown is the relative abundance [X/Fe].   The original composition  of the PopIII-SNII ejecta is shown by the plus signs in all of the panels.
    These elements are   identified 
   by an asterix next to the their  names in all of the panels as their abundances vary with $t_{\rm delay}$.
   
    For $t_{\rm delay}> t_{\rm no-spall}\sim $ 23-24 days
    (not shown here) no spallation occurs and the original elements  (identified with an asterix) fall exactly on the plus signs.
  For   $ t_{\rm outer}\sim 8.3\ {\rm days}  < t_{\rm delay} < t_{\rm no-spall}$,  
    there is minimal interaction in the inner layers and most spallation products are light elements from
    O- and C-spallation.  The original relative
    abundance of the PopIII-SNII elements is almost conserved except for
    some spallation in the oxygen layer which leads to the  formation of B, Be and Li.
    For intermediate $t_{\rm delay}$ (i.e. $t_{\rm inner}\sim 6.3\ {\rm days}  < t_{\rm delay} < t_{\rm outer}$), spallation occurs
    in almost all of the layers.     As $t_{\rm delay}$ gets shorter  spallation efficiency start to decrease in the outer layers 
    and eventually for $t_{\rm delay}<   \sim $ 6 days spallation starts to become more prominent in the
    innermost layers.    For even shorter $t_{\rm delay}$ (top panels), most of the QN neutrons
   are used up in the $^{56}$Ni layer thus shielding all of the overlaying layers.  
    Effectively, the original PopIII-SNII sub-Ni    elements (S, Si, Mg, Ne, O, C) present in the PopIII-SNII  ejecta would be preserved 
 in dsQNe with very short delays.  In addition these elements will be produced as by-products of spallation
  in the $^{56}$Ni layer which further increases   their overall [X/Fe] values  for $t_{\rm delay} < 3$ days.
   The weak-dependency of multiplicity on the neutrons' energy
explains the nearly similar final  abundances for  the $E_{\rm QN}=$ 5 GeV and the $E_{\rm QN}=$ 10 GeV simulations.

  \subsubsection{$\alpha$-elements}
  \label{sec:alphas}
  
  The so-called $\alpha$-elements (O, Ne, Mg, Si, S, Ca, Ti) in very metal-poor stars
 are  reported to be overabundant relative to iron at low metallicity (e.g. McWilliam 1997;  Aoki et al. 2007) 
  with  $[\alpha/Fe]  > [\alpha/Fe]_{\odot}$ where 
 $[\alpha/Fe] = \frac{1}{4}( [Mg/Fe]  + [Si/Fe]  + [Ca/Fe]  + [Ti/Fe])$ (the average abundance of Mg, Si, Ca and Ti). These enhancements increase linearly with decreasing metallicity.

\begin{sidewaysfigure}
%\begin{figure*}
\centering
\includegraphics[scale=1.2]{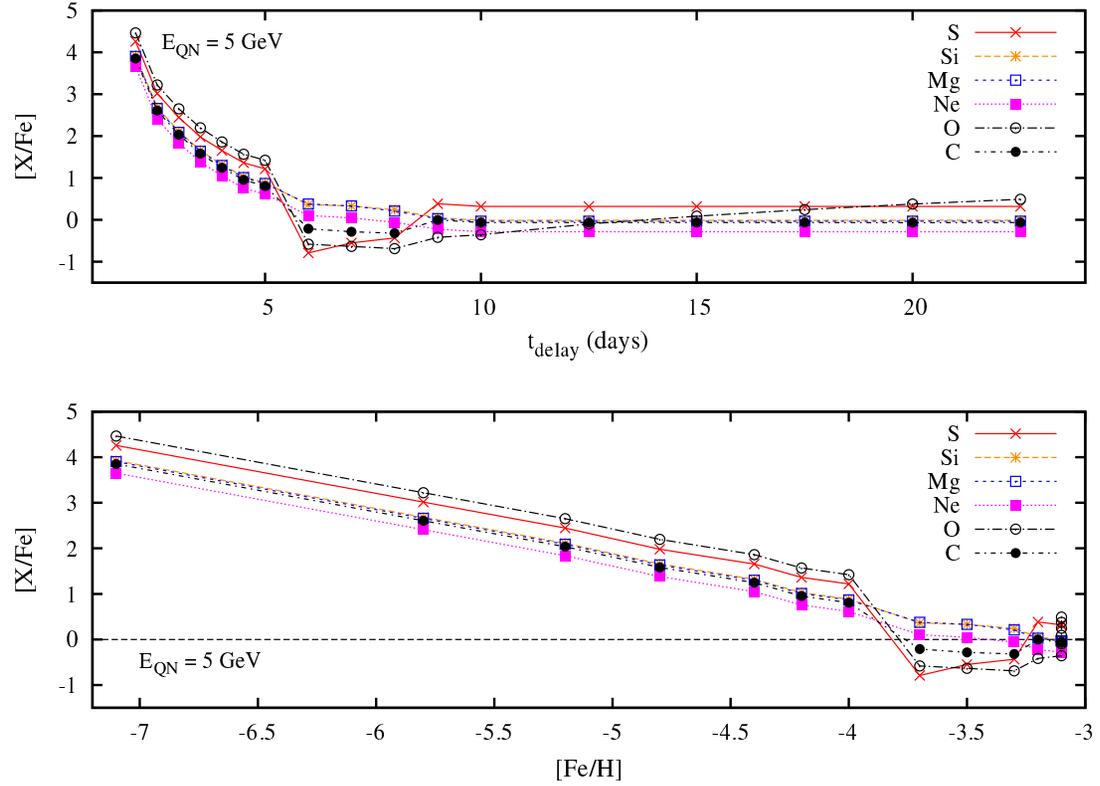} 
\caption{Relative abundances [X/Fe] versus
$t_{\rm delay}$ (top panel) and versus the corresponding ${\rm [Fe/H]} $ (bottom panel) for
 $^{32}$S, $^{28}$Si, $^{24}$Mg, $^{20}$Ne, $^{16}$O and $^{12}$C with initial relative abundances [X/Fe]$_{\rm X,SN}$=0.32, -0.01, -0.03, -0.28, 0.53, -0.09, respectively.  
These elements represent the original composition of the PopIII-SNII  ejecta (i.e. the target material) in our model.
  Here, $E_{\rm QN}=5$ GeV.  The horizontal dashed line in the bottom panel corresponds to [X/Fe]=0.
}
 \label{fig:alpha-elements-5GeV}
% \end{figure*}
 \end{sidewaysfigure}

\begin{sidewaysfigure}
% \begin{figure*}
\centering
 \includegraphics[scale=1.2]{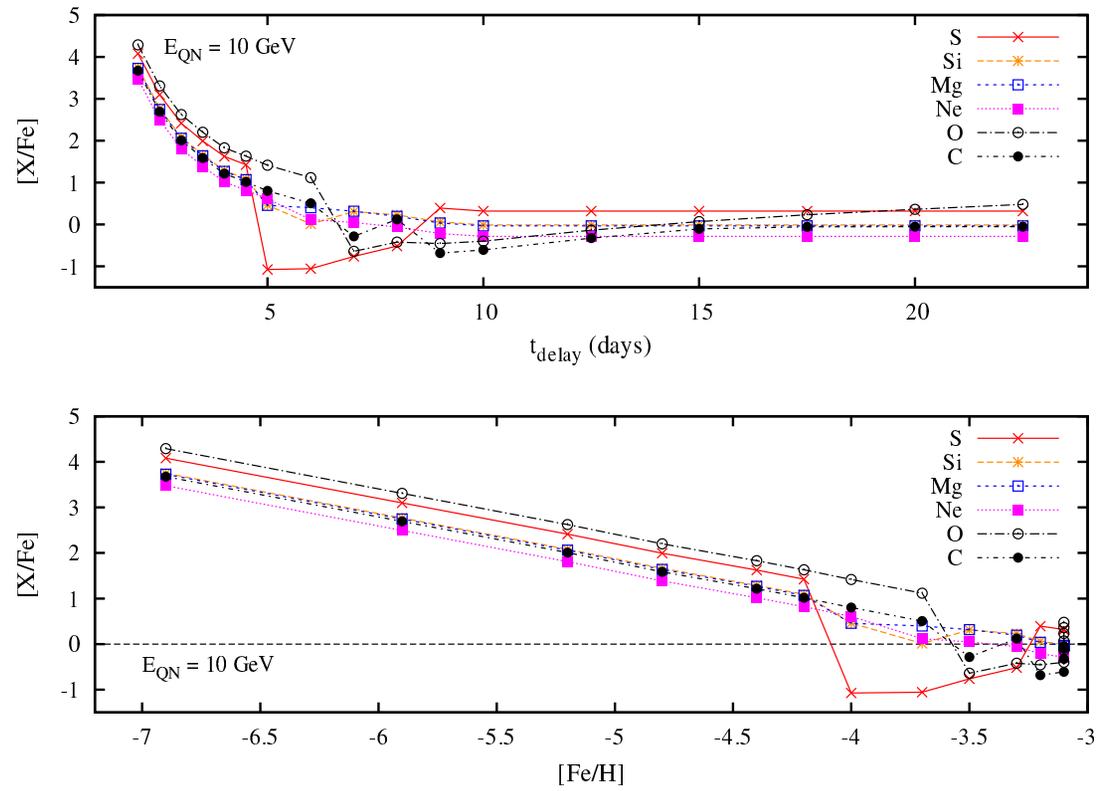} 
 \caption{Same as in figure \ref{fig:alpha-elements-5GeV} but for $E_{\rm QN}=10$ GeV.}
 \label{fig:alpha-elements-10GeV}
%\end{figure*}
 \end{sidewaysfigure}

  Figure \ref{fig:alpha-elements-5GeV}  shows the relative abundances of the original PopIII-SNII elements with
respect to $t_{\rm delay}$ (top panel) and metallicity (lower panel) for simulations with $E_{\rm QN}=5$ GeV.
 Figure \ref{fig:alpha-elements-10GeV} shows
the results  for the $E_{\rm QN}=10$ GeV simulations.   For $t_{\rm delay} > t_{\rm outer}\sim 8.3$  days, [X/Fe]
remain close to the original values except for oxygen which experiences some  spallation and thus
depletion up to $t_{\rm delay}\sim 15$ days since it is by far the most abundant PopIII-SNII element.
For $t_{\rm inner} < t_{\rm delay} < t_{\rm outer}$,
  spallation occurs in most of the PopIII-SNII layers leading to an overall reduction in the abundances
  of the original elements.   This leads to a dip in the [X/Fe] values  in Figure \ref{fig:alpha-elements-5GeV} 
   and in Figure \ref{fig:alpha-elements-10GeV}  for $-4 <$ [Fe/H] $<-3$.
    A plateau-like behaviour is noticeable in the dip for all of the PopIII-SNII elements.     
      The [Si/Fe] and [Mg/Fe] levels in the dip region,  combined with the copious
 production of Ti and Ca from Ni-spallation (see Figure \ref{fig:X-over-Fe})
 lead to $[\alpha/Fe]  >  [\alpha/Fe]_{\odot}$ in our model and in particular  
 for $t_{\rm delay} < t_{\rm outer}\sim 8.3$ days (i.e. ${\rm [Fe/H]}  < -3$) since spallation is more pronounced in the
 $^{56}$Ni layer than in the overlaying layers.  
  
  An increase of $[X/Fe]$ is evident
for most elements when the metallicity decreases below $\sim -4$; i.e. when $t_{\rm delay}$ decreases below roughly 5 to 6 days. 
These layers are shielded at the expense of $^{56}$Ni
  depletion explaining the increase in [X/Fe]  much above the initial values.
 In general,  and for short $t_{\rm delay}$ in our model, the more abundant a specific element
 is in the PopIII-SNII shocked ejecta, prior to spallation by the QN neutrons, the more it will appear abundant  following the destruction
 of $^{56}$Ni (i.e. the reduction of the final  iron abundance in the swept up cloud).  Since O and C are the two-most abundant elements
 in the PopIII-SNII ejecta, it is no surprise
   that [C/Fe] and [O/Fe] are very enhanced for very short delays (see below for the special of case of $^{32}$S). 
These enhancement  would be reflected 
in the composition of  the gas cloud swept up by the SN+QN ejecta out of which the Pop. II low-mass stars   would form.

    \subsubsection{Sulfur}
  \label{sec:sulfur}

   Figures \ref{fig:alpha-elements-5GeV} and  \ref{fig:alpha-elements-10GeV}  show that Sulfur is more depleted (with [S/Fe]  showing  the lowest plateau in the dip)  than the
  other original elements in the PopIII-SNII ejecta for $6.3\ {\rm days} < t_{\rm delay} < 8.3\ {\rm days}$ (i.e. $-4 <$ [Fe/H] $<-3$)
  due to its close proximity to the Nickel layer.   Even for  very short delays, because of the
  non-uniform distribution in multiplicity $\zeta$, some spallated neutrons in the
  $^{56}$Ni manage to make it to the overlaying $^{32}$S layer.    This explains why 
   the $^{32}$S plateau in the dip  extends to metallicity as low as ${\rm [Fe/H]} \sim -4$. 
      Only for $t_{\rm delay} <$ 5 days (i.e. [Fe/H]$< -4$) does
  it become completely shielded and starts to  follow the behavior of the other elements by increasing with decreasing [Fe/H].
   $^{32}$S has another distinctive feature:  [S/Fe] exceeds [C/Fe] at very low metallicity despite the
    higher  initial abundance of C as compared to S in the PopIII-SNII ejecta.  It turns out that 
 Sulfur is among the main spallation products from  Ni-spallation  for $t_{\rm delay} < 5$ days (i.e. ${\rm [Fe/H]} \sim -4$).
     This produced S adds to the original S   which explains why [S/Fe] exceeds [C/Fe] for our fiducial values.
      The trend  of [S/Fe] versus [Fe/H]  in extremely metal-poor stars  awaits  monitoring and observations  
 before its dependence on metallicity are known with higher accurcay (e.g. Nissen et al. 2007; Spite et al. 2011 and 
 references therein).   Only then can our model be put to test.
  
  \subsubsection{Comparing to the average star}
  \label{sec:averagestar}

\begin{figure*}
\begin{center}
\includegraphics[height=0.55\textwidth, width=0.75\textwidth]{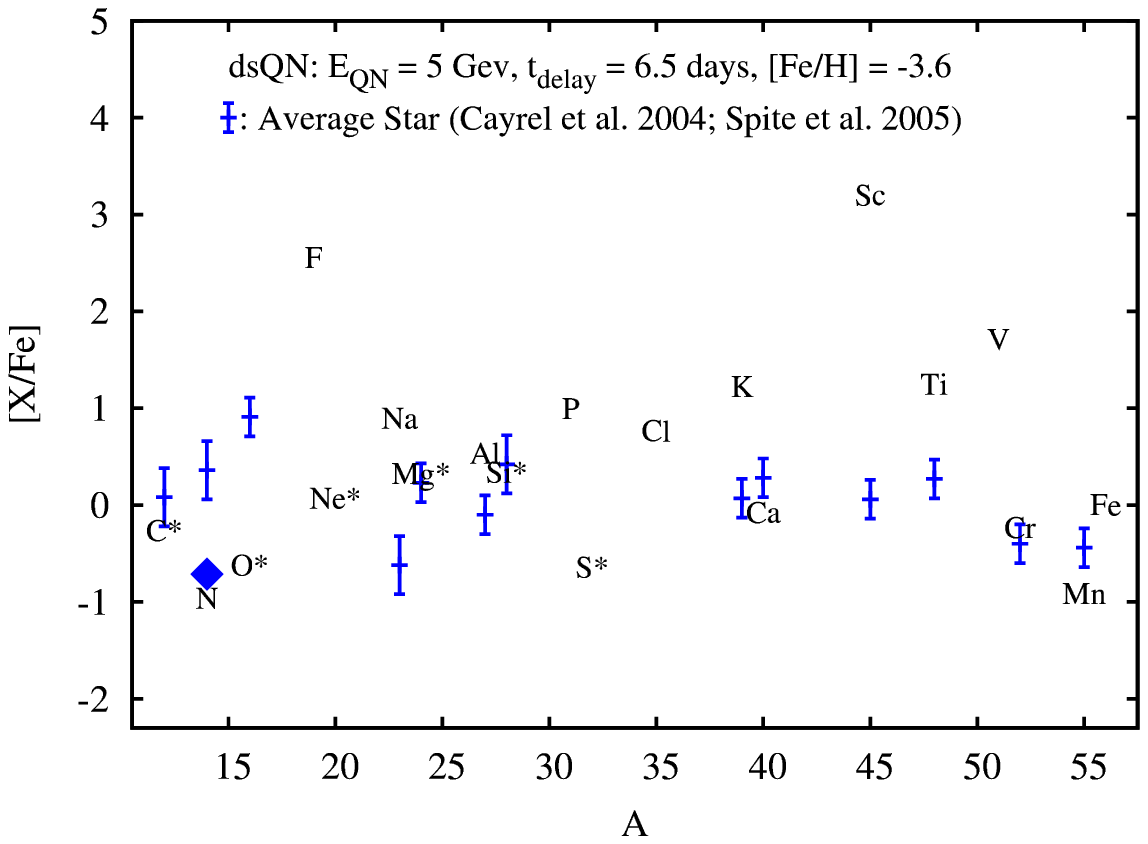}\\
\includegraphics[height=0.55\textwidth, width=0.75\textwidth]{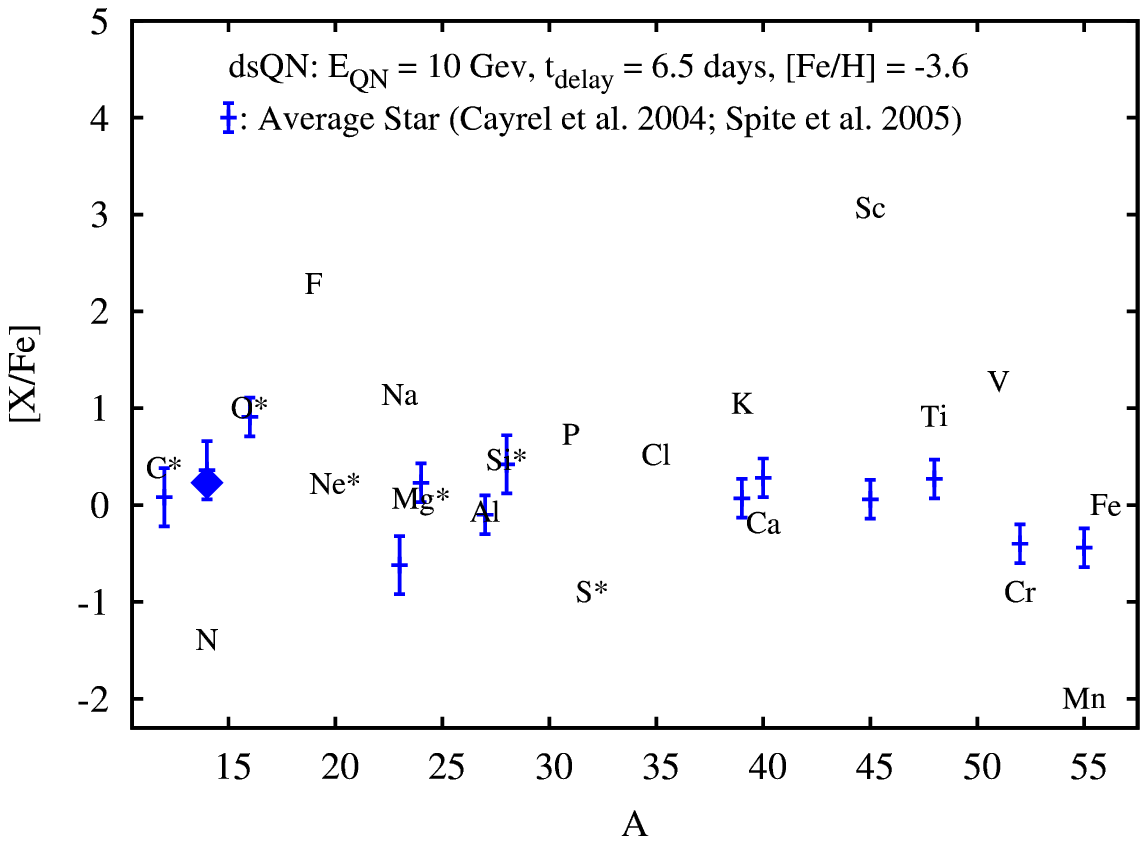}
\caption{Relative abundances  [X/Fe] of sub-Fe spallation products (elements identified by their names) 
versus mass number $A$
from simulations with $t_{\rm delay}=6.5$ days; this resulted in ${\rm [Fe/H]} \simeq -3.6$.  
The CN enhanced nitrogen is shown by the filled diamond. 
 Measured  abundances  of an average halo star  (Cayrel et al. 2004; Spite et al. 2005) 
 are also shown as plus ($+$) signs, with corresponding uncertainties,  
 for a comparison.  
  Top and bottom panels are for QN neutrons with $E_{\rm QN}=5$ GeV and 10 GeV, respectively. 
}
 \label{fig:average-star}
 \end{center}
\end{figure*}

  Figure \ref{fig:average-star} is a comparison of yields from a typical  dsQN 
experiencing spallation in the inner and outer layers   (i.e. in the  $t_{\rm inner} < t_{\rm delay} < t_{\rm outer}$
regime which corresponds to $-4 <$ [Fe/H] $<-3$ in our model); we chose $t_{\rm delay}= 6.5$ days. Our simulation results are compared to 
abundance pattern shared by stars in the metallicity range $- 4.0 <$ [Fe/H] $< -3.0$
with a normal [C/Fe] ratio (Cayrel et al. 2004; Spite et al. 2005; see also Figure 1 in Limongi\&Chieffi 2012).
As can be seen from Figure \ref{fig:average-star}, both the $E_{\rm QN}=$ 10 GeV and  $E_{\rm QN}=$ 5 GeV simulations are
successful in reproducing the observed abundances.  Note the the high abundances of Fluorine and Scandium which are spallated in the innermost layers in our model; these two rare elements are natural products of   spallation 
(see  \S 6.2 in Ouyed (2013).  The $E_{\rm QN}=$ 5 GeV simulations underestimate
the [C/Fe] and [O/Fe] abundances but this is easily remedied by increasing the original  carbon and oxygen content
in the PopIII-SNII ejecta to reasonable amounts.

\subsection{Nitrogen in our model}
\label{sec:nitrogen}

As shown in Ouyed (2013; see \S 3.3 in that paper), $^{14}$N forms by spallation in
 the $A>14$ layers in the PopIII-SNII ejecta. However, we find that  
$^{14}$N  can also be  enhanced by the injection of protons from spallation into the C-rich and O-rich outer
      layers if temperatures in the PopIII-SN ejecta
      exceeds  $\sim 1.6\times 10^7$ K when it is hit by the QN ejecta.  The carbon (and perhaps oxygen) could subsequently be processed into $^{14}$N via the CN(O) cycle (Arnett 1996).
For an adiabatic expansion of the preceding SN shell (i.e. $T_{\rm env.}\propto t^{-2}$), the $T_{\rm env.} > 1.6\times 10^7$ K necessary condition  for CN (proton capture) process to take place yields
\begin{equation}
 \label{eq:tLi}
t_{\rm delay} < \sim 5\ {\rm days}\ \frac{R_{\rm prog., 100}T_{\rm env., 10}^{1/2}}{v_{\rm sn, 5000}}\ ,
\end{equation}
where $R_{\rm prog., 100}$ is the initial radius of the progenitor in units of $100R_{\odot}$ ($R_{\odot}$ is the solar radius)  and $T_{\rm env., 10}$ the
 initial temperature of the PopIII-SNII envelope in units of $10^{10}$ K.
For our fiducial parameters, the above translates to (see Figure 1 in Ouyed 2013)
 \begin{equation}
\left[ \frac{Fe}{H}\right] < -4  \ .
 \end{equation}
 We find  that the  CN(O) processing of a small percentage ($<$ 10\%) of the carbon and oxygen into $^{14}$N 
 is enough to account for the observed [N/Fe] values in an average star (in Figure  \ref{fig:average-star}
  the CN enhanced nitrogen is identified with a filled diamond).
 The  $^{13}$C  abundance should also be enhanced if  CN equilibrium is established in which case 
    the $^{12}$C/$^{13}$C ratios  should be of the order of a few (see however Caughlan \& Fowler 1972).
    In our model, $^{14}$N production from CN cycling could occur  at higher [Fe/H] when taking into account envelope reheating
 by the QN shock (following impact on the SN ejecta).
However, we caution that our model is more detailed than the standard CN cycle and would need further investigations for better estimates of $^{14}$N enhancement and the   expected $^{12}$C/$^{13}$C values.
    
\section{Lithium synthesis  in our model}
\label{sec:lithium}

The $^7$Li produced by spallation on a target $A_{\rm T}$ in the ejecta of the PopIII-SNII ejecta  is given
 by eq.(14) in Ouyed (2013):
 \begin{equation}
 \label{eq:ALi}
 A(Li) \simeq \log{\eta_7^{\rm A_T}} + \log{ \frac{ M_{\rm A_T, SN}/1.5 M_{\odot} } { M_{\rm sw}/10^5 M_{\odot} } }  + 6.33\ ,
 \end{equation}  
 where  $\eta_{\rm ^7}^{A_{\rm T}} = M_{\rm ^7Li}/M_{\rm A_{\rm T, SN}}$ is the normalized mass yields of $^7$Li 
following spallation on target $A_{\rm T}$ with initial mass $M_{\rm A_T, SN}$; the
$1.5M_{\odot}$ is representative of our fiducial value for the oxygen content  in mass 
in the PopIII-SNII ejecta. The final $^7$Li abundance is a combination of $^7$Li produced from spallation on
each of the targets present in the PopIII-SNII ejecta.

\begin{figure*}
 \begin{center}
\begin{tabular}{cc}
\includegraphics[height=0.35\textwidth,width=0.55\textwidth,]{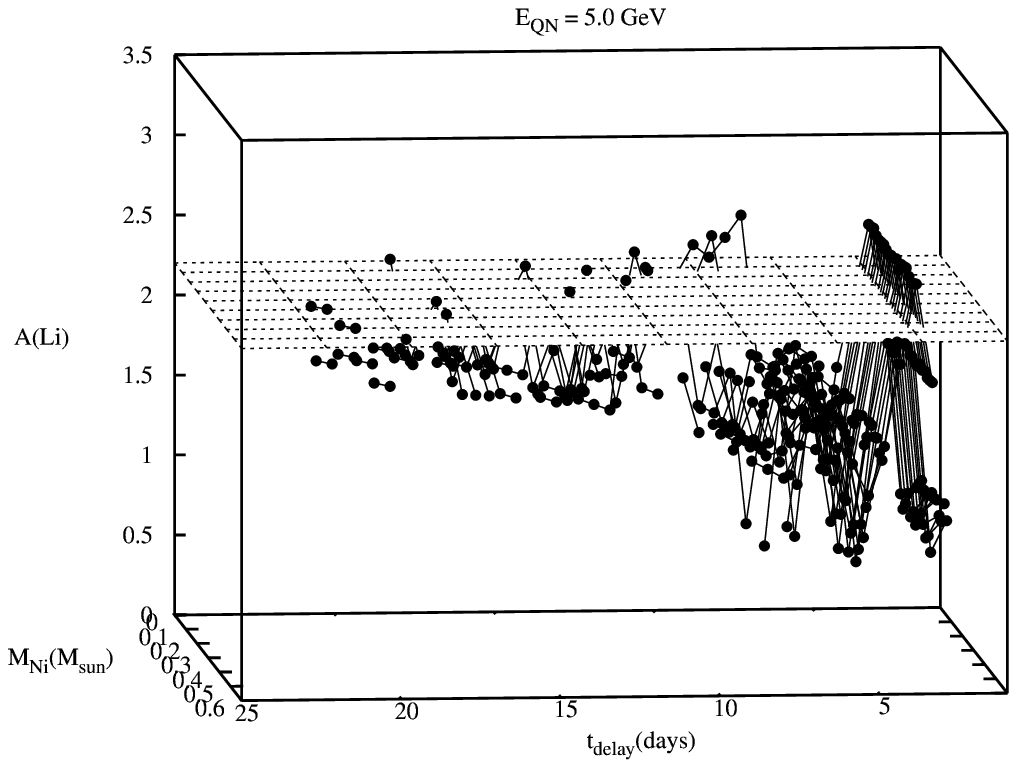} &  \includegraphics[height=0.35\textwidth,width=0.55\textwidth]{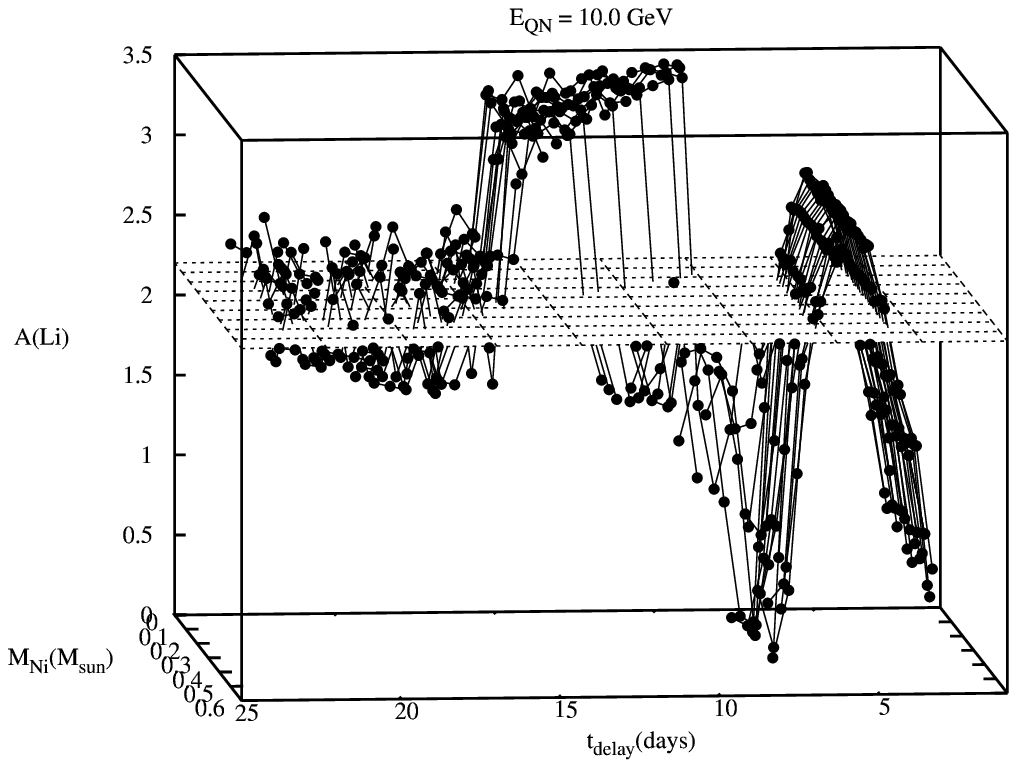} \\
\includegraphics[height=0.35\textwidth,width=0.55\textwidth]{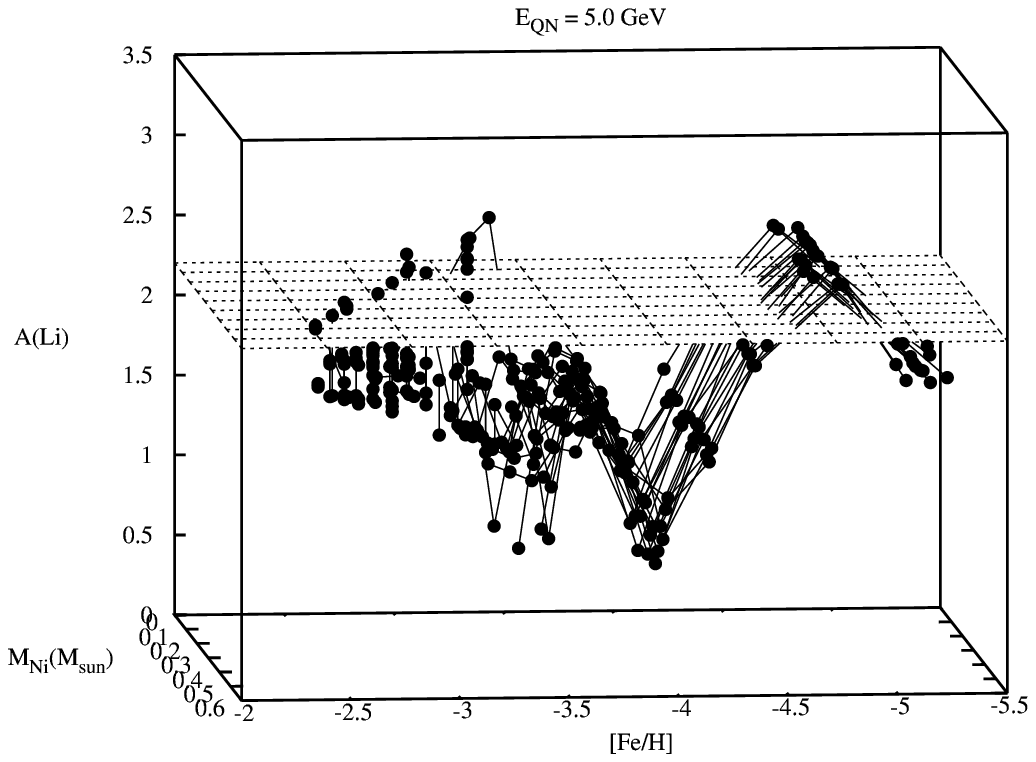} &  \includegraphics[height=0.35\textwidth,width=0.55\textwidth]{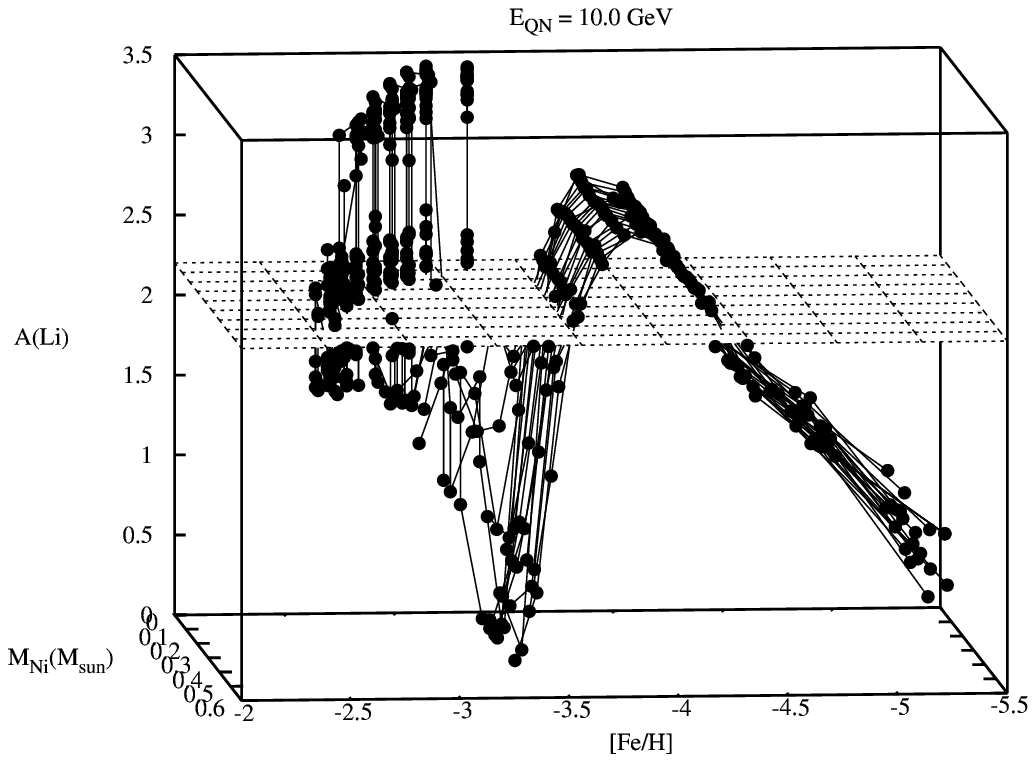} \\
\end{tabular}
\caption{
Top panels: $A(Li)$ versus time delay   ($ 2\ {\rm days}  \le  t_{\rm delay} \le 30\ {\rm days} $) and 
  PopIII-SNII initial $^{56}$Ni mass ($0.1 M_{\odot} \le M_{\rm Ni, SN} \le 0.5M_{\odot}$).  Bottom panels:  The corresponding $A(Li)$ versus ${\rm [Fe/H]} $ ($ -7.5 < {\rm [Fe/H]}  < -2.5$; see \ref{sec:tdelay-FeH}).     Left-hand and right-hand panels are for $E_{\rm QN}=5$ GeV and 10 GeV, respectively.
  The horizontal plane corresponds to $A(Li)=2.2$.
}
 \label{fig:3D-Li}
  \end{center}
\end{figure*}

 Shown in Figure \ref{fig:3D-Li} is the $^7$Li abundances resulting from spallation by $E_{\rm QN}=5$ GeV (left panels)
  and $E_{\rm QN}=10$ GeV (right panels) QN neutrons on a typical PopIII-SNII ejecta in our model.  
  The top panels show $A(Li)$ 
  from  simulations spanning a range in $^{56}$Ni content of the PopIII-SII ejecta ($0.05 M_{\odot} \le M_{\rm Ni, SN} \le  0.5 M_{\odot}$)
  and a range in $t_{\rm delay}$ (2 days $\le t_{\rm delay} \le $ 24 days). The bottom panels
  show the corresponding metallicity [Fe/H];  as explained in \ref{sec:tdelay-FeH} there is a direct connection
  between $t_{\rm delay}$ and [Fe/H] in our model.
  For $t_{\rm delay} >  $ 24 days the QN neutrons traverse the
   PopIII-SNII ejecta with minimal interaction; i.e. no spallation occurs and the PopIII-SNII ejecta  abundances would not be modified or
   affected (i.e. $\eta_{\rm A}^{\rm A_T}=1$). While $^{56}$Ni content was varied, the other initial  abundances in mass 
    in the PopIII-SNII were  set to their fiducial values: $M_{\rm S, SN}=M_{\rm Si, SN}=M_{\rm Mg, SN}=M_{\rm Ne, SN}= 0.05M_{\odot}$, $M_{\rm O, SN}=1.5M_{\odot}$ and $M_{\rm C, SN}= 0.15M_{\odot}$.  We   adopt  a $10^5M_{\odot}$ of pristine
  cloud (the cradle of the Pop. II low-mass stars) swept up by the SN+QN ejecta (e.g. Shigeyama\&Tsujimoto 1998).

\begin{figure*}
\begin{center}
\includegraphics[height=1.0\textwidth,width=1.0\textwidth]{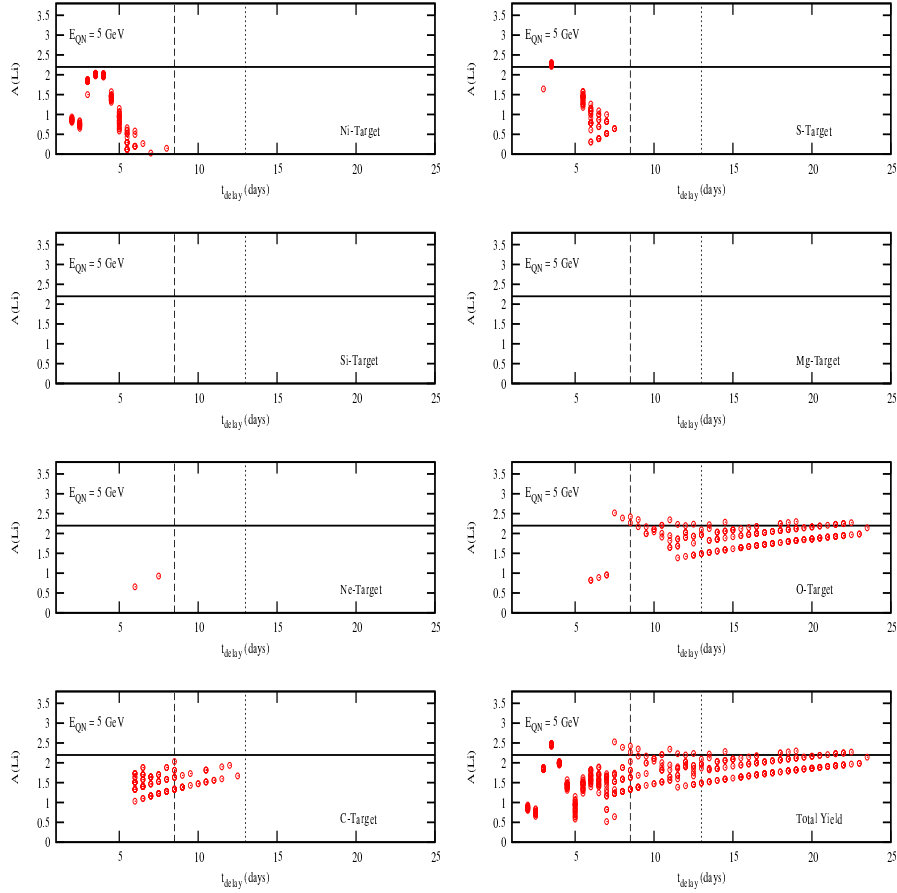}
\caption{The open circles show $A(Li)$ versus $t_{\rm delay}$  per  spallation  target  (Ni, S, Si, Mg, Ne, O and C) for the simulations
  shown in the left panels in figure \ref{fig:3D-Li}  (i.e. for $E_{\rm QN}=$ 5 GeV).
   The "Total Yield" panel  shows  $A(Li)$ from all targets combined.  The Horizontal lines  corresponds to  $A(Li) = 2.2$. 
    The vertical dashed line shows $t_{\rm outer}\sim 8.3$ days while the dotted vertical line  is at 13 days
    which defines the $\Delta R_{\rm C}= \lambda_{\rm sp., C}$ limit (i.e. $N_{\rm mfp, C}=1$) above which no spallation
    occurs in the C layer.}
    \label{fig:targets-days-5GeV}
 \end{center}
 \end{figure*}

\begin{figure*}
\begin{center}
\includegraphics[height=1.0\textwidth,width=1.0\textwidth]{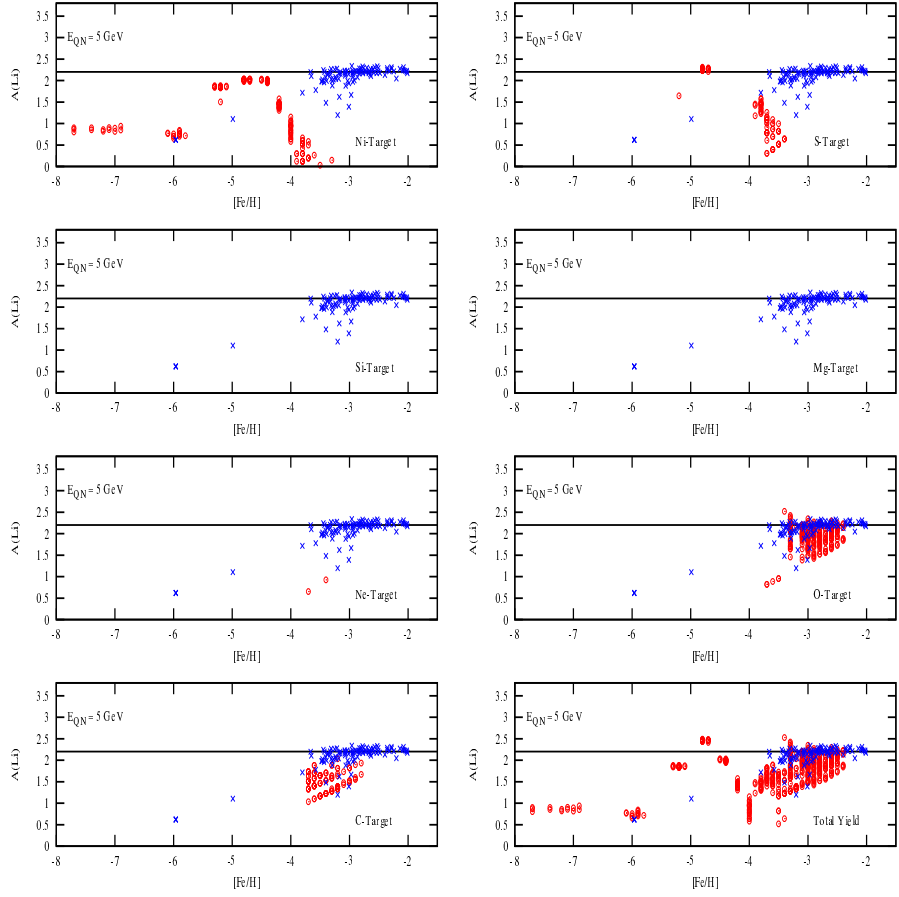}
\caption{The open circles show $A(Li)$ versus [Fe/H]   per  spallation  target  (Ni, S, Si, Mg, Ne, O and C) for the simulations
  shown in the left panels in figure \ref{fig:3D-Li}  (i.e. for $E_{\rm QN}=$ 5 GeV).
   The crosses are measured $^7$Li abundances  in halo and turn-off  stars from Sbordone et al. (2012;   see their figure 1 for references).
   }
 \label{fig:targets-FeH-5GeV}
 \end{center}
 \end{figure*}

\begin{figure*}
\begin{center}
\includegraphics[height=1.0\textwidth,width=1.0\textwidth]{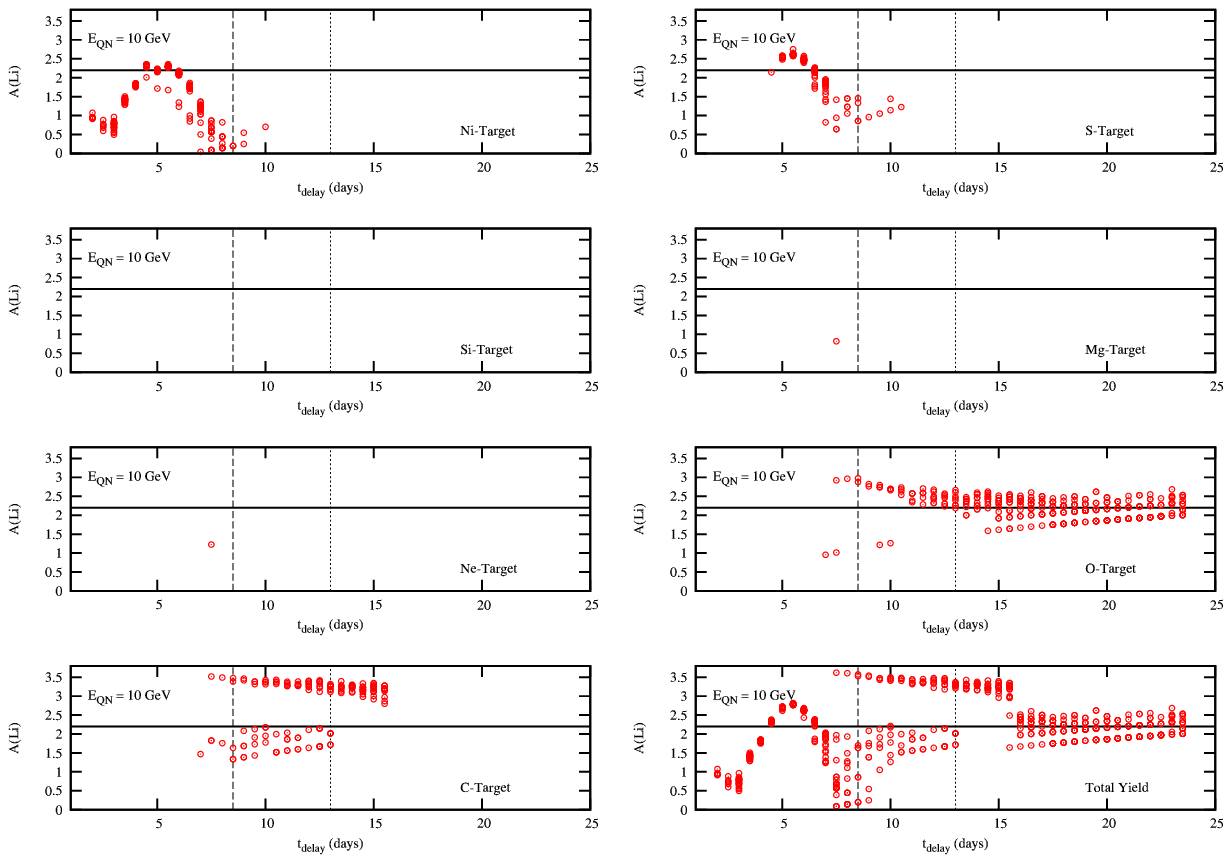}
\caption{Same as in figure \ref{fig:targets-days-5GeV} but for the $E_{\rm QN}=$ 10 GeV simulations.}
 \label{fig:targets-days-10GeV}
  \end{center}
 \end{figure*}

\begin{figure*}
\begin{center}
 \includegraphics[height=1.0\textwidth,width=1.0\textwidth]{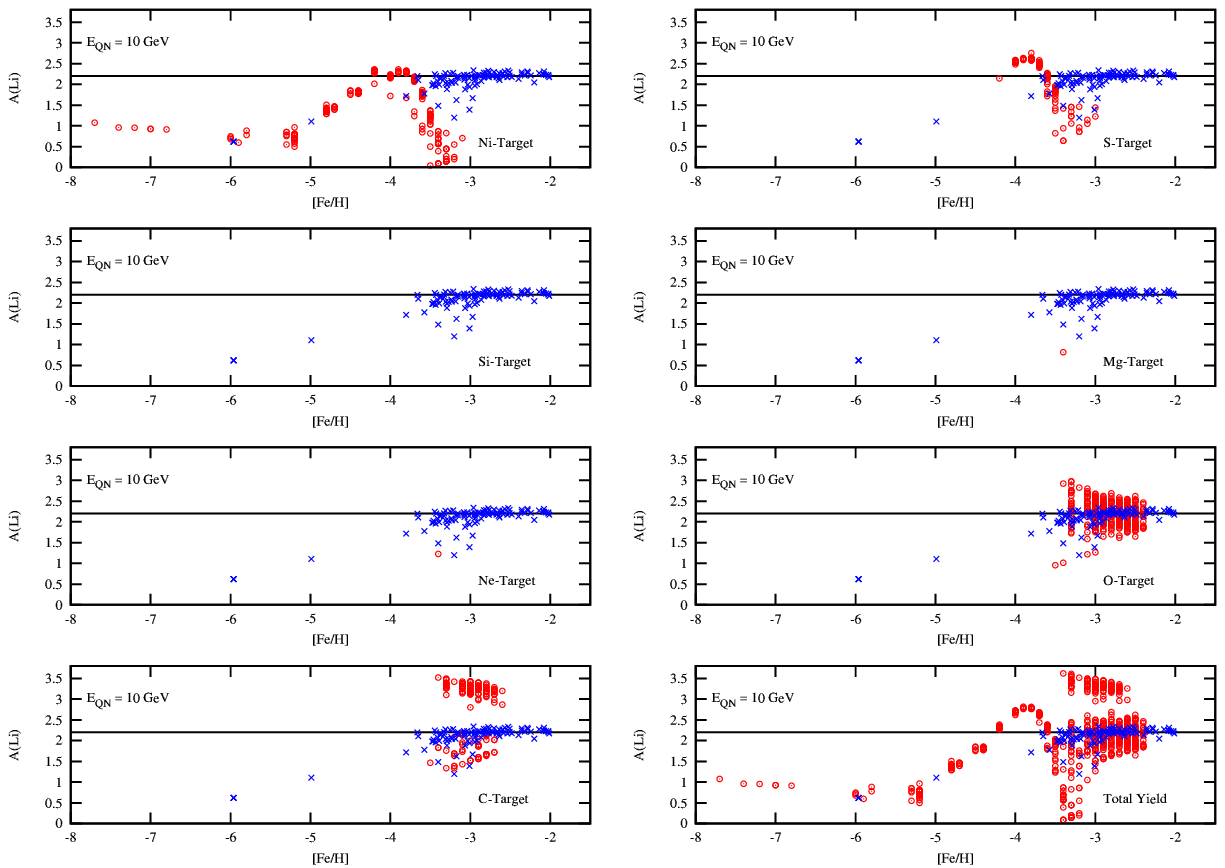}
 \caption{Same as in figure \ref{fig:targets-FeH-5GeV} but for the $E_{\rm QN}=$ 10 GeV simulations.}
 \label{fig:targets-FeH-10GeV}
 \end{center}
\end{figure*}

   The different panels in Figure \ref{fig:targets-days-5GeV}  (Figure \ref{fig:targets-days-10GeV} ) show the resulting $A(Li)$ versus $t_{\rm delay}$ for each target
   for the $E_{\rm QN}=5$ GeV ($E_{\rm QN}=10$ GeV) simulations.  The corresponding $A(Li)$ versus [Fe/H] is shown in Figure \ref{fig:targets-FeH-5GeV} (Figure \ref{fig:targets-FeH-10GeV}).
   The lower right panel in  Figure \ref{fig:targets-days-5GeV}  (Figure \ref{fig:targets-days-10GeV}) shows  the combined $A(Li)$ versus  $t_{\rm delay}$  
     which  effectively corresponds  to the projection of  the upper left  (right) panel in Figure   \ref{fig:3D-Li} onto the $A(Li)$-$t_{\rm delay}$ plane. 
     Similarly,  the lower right panel in  Figure \ref{fig:targets-FeH-5GeV}  (Figure \ref{fig:targets-FeH-10GeV}) shows  the combined $A(Li)$ versus  [Fe/H]  
     which  effectively corresponds  to the projection of  the lower left  (right) panel in Figure   \ref{fig:3D-Li} onto the $A(Li)$-[Fe/H] plane. 
    One can see from these panels  that $^7$Li is produced in the
     $^{56}$Ni and $^{32}$S layers when $t_{\rm delay}< t_{\rm outer}$  and in the
      oxygen and carbon layers for longer delays ($t_{\rm outer} \sim $ 8.3 days is shown by  the vertical dashed 
     lines in the figures).  No  $^7$Li is spallated in the intermediate layers (Si, Mg, Ne) and in general $^7$Li 
      is mainly produced either in the innermost layers or in the outermost CO  layers.  The vertical doted lines defines  the $N_{\rm mfp, C} = \Delta R_{\rm C}/\lambda_{\rm sp., C} =1$ boundary 
      which defines a timescale above which the spallation mfp exceeds the thickness
      of the C-layer; i.e. no C-spallation occurs for longer time delays even if the neutrons exiting the O-layer have energies exceeding 
      $E_{\rm sp., C}\sim 0.42$ GeV which is above the
      critical value for C-spallation to ensue.

     There is a gap in the $^7$Li abundance (i.e. a drop in the $A(Li)$ value)
  in the intermediate regime, $t_{\rm inner} < t_{\rm delay} < t_{\rm  Ni}$ (which corresponds to $- 4<$ [Fe/H] $<-3$)
   since most neutrons end up being used for spallation of $A>7$ elements in the inner layers.
   Thus,  for Pop. II low-mass stars  formed from clouds swept up by dsQNe with $t_{\rm delay}< $ 8.3 days
        (i.e. for [Fe/H]$<-3$), the resulting $^7$Li abundance would show a larger scatter mainly because
         in these dsQNe $^7$Li is not the main spallation product and the neutrons are used up to make many
         other spallation products.   At this point, 
                  we must note   that the $^7$Li abundance in dsQNe leading to [Fe/H]$<-3$ 
                     (i.e. in   $t_{\rm delay}  < \sim$ 10  days 
           dsQNe) should be considered
  upper limits in our model.  For such dsQNe,  
   the SN ejecta is still hot ($T> 2.5\times 10^6$ K)  when it is spallated.     Combined with the
   copious production of protons (see Tables \ref{tab:ni-spallation} and  \ref{tab:o-spallation}), we expect  most of the spallated  $^7$Li  to be destroyed  by reactions with the  protons  or at least become 
   reduced   (see \S 5 in Ouyed 2013).

   \subsection{The high-scatter, $A(Li)\sim 2.2$, plateau}
  \label{sec:o-plateau}
   
   Two ``plateaus" can be seen in the top right panel in Figure \ref{fig:3D-Li} (i.e. for the $E_{\rm QN}=10$ GeV simulation). One
 resulting from O-spallation with $1.8 <A(Li) < 2.5$ and a second one from C-spallation
with $2.5 <A(Li) < 3.5$.   These two plateaus can also be seen in panels labeled ``C-Target"
and ``O-Target" in Figures  \ref{fig:targets-days-5GeV},  \ref{fig:targets-FeH-5GeV},\ref{fig:targets-days-10GeV},   \ref{fig:targets-FeH-10GeV}.
 Both   occur for $t_{\rm delay}> t_{\rm outer}\sim 8.3$ days which means that the QN neutrons
         proceed directly into the outer layers with minimal interaction with the Ni, S, Si, Mg and Ne layers.
          Spallation in the O-layer leads to  smaller $^7$Li abundance and more scatter in the $^7$Li plateau since the neutrons spallate 
   more products  than in the case of spallation in the C-layer.      
         
    We first focus on the $A(Li)\sim 2.2$ plateau uncovered in our simulations (the  $A(Li)= 2.2$ value
     is indicated by the horizontal planes in Figure \ref{fig:3D-Li} and by the horizontal lines in Figures \ref{fig:targets-days-5GeV},
    \ref{fig:targets-FeH-5GeV}, \ref{fig:targets-days-10GeV}, \ref{fig:targets-FeH-10GeV}). This plateau  
 is populated by $^7$Li spallated from oxygen for dsQNe with $t_{\rm delay}> 8.3$ days.
   As can be seen from Figure \ref{fig:3D-Li}, while the   $E_{\rm QN}=5$ GeV simulations yield $A(Li)$ approaching the 2.2 value from the lower
           bound, barely exceeding it, the $E_{\rm QN}=10$ GeV simulations yield $^7$Li
           abundances  oscillating around  $A(Li)\sim 2.2$ (with  $1.8 < $A(Li)$ < 2.5$).   
      To understand this interesting finding from our simulations we first note that
   in each sub-layer $i$ (i.e. for each mfp, $\lambda_{\rm sp., O}$)  of the O-layer,  the distribution of the atomic weights of the spallation products  is close to 
          a normal distribution peaking at $A_{\rm peak, i}\sim A_{\rm T}- \zeta_{\rm av., i}$ where $\zeta_{\rm av., i}$
          is the average multiplicity for sub-layer $i$ (see Table \ref{tab:o-spallation}). For each sub-layer the   corresponding amount  in mass of spallated $^7$Li can be estimated from the corresponding  distribution.

          Since the highest $\zeta_{\rm av.}$ is induced in the first sub-layer this
                    leads to a peak spallation product ($A_{\rm peak}= 16-\zeta_{\rm av.}$)
            closest  to $^7$Li.  Subsequent sub-layers would yield peak spallation products getting
            closer to the target's $A_{\rm T}$ (here $A_{\rm T}=16$) thus reducing the $^7$Li contribution to the overall 
            distribution of the spallated elements and thus a reduction in the final amount of $^7$Li spallated. However, for subsequent sub-layers 
            the multiplicity is higher and more neutrons$+$protons induce spallation.  
          The final abundances of the spallation products is a combination of these distributions (see e.g. Figure 1 in Ouyed et al. 2011) which yields $M_{\rm Li}\sim \alpha_{7} \zeta_{\rm net} M_{\rm QN}\equiv \beta_{7} M_{\rm QN}$ where $\beta_{7} =  \alpha_{7} \zeta_{\rm net}$.
          Since  the  $^7$Li contribution given by the distributions is roughly $\alpha_{7} \sim 0.01$ (of the order of $\sim 1\%$)  and $\zeta_{\rm net}\sim 10$ (see e.g. Table \ref{tab:o-spallation}) this yields $ \beta_{7} \sim 0.1$
             The corresponding $^7$Li abundance $A(Li) = \log{N_{\rm ^7Li}/N_{\rm H}} + 12$ is then roughly estimated to be
          \begin{equation} 
          A(Li) \sim 2.15 + 
          \log{\left( \beta_{7, 0.1} \frac{M_{\rm QN}/10^{-3} M_{\odot}}{M_{\rm sw}/10^5 M_{\odot}}\right)} \ ,
          \end{equation}
          where we made use of $N_{\rm ^7Li}=M_{\rm ^7Li}/7 m_{\rm H}$ and $N_{\rm H}= M_{\rm sw}/m_{\rm H}$. 
        The amount of $^7$Li spallated is roughly equivalent to a  normalized  mass yield 
        $\eta_7^{16} \sim 6.7\times 10^{-5}$ or  $\log{ \eta_7^{16} }\sim -4.17$ for our fiducial values; i.e.  $A(Li)\sim 2.15$ from using eq.(\ref{eq:ALi}).
         
         The weak dependence of the multiplicity on energy leads to similar distributions of spallation products in the oxygen layer 
           for the range in $E_{\rm QN}$ expected  in QNe explosions. This translates to similar $\zeta_{\rm net}$ and $\alpha_7$  (i.e $\beta_7$) values which  explains why the $E_{\rm QN}=5$ GeV also hovers around the same $A(Li)$ value and effectively 
           implies that $A(Li)$ is mainly dependent on $M_{\rm QN}$ and $M_{\rm sw}$.  In fact by scaling $M_{\rm sw}$ with
           $M_{\rm QN}$ (i.e. with the kinetic energy, $E_{\rm QN}^{\rm KE}$, of the QN ejecta) we always recover an $A(Li)$
           close to Spite values when $t_{\rm delay}> t_{\rm outer}$; i.e. when spallation occurs mainly in the outer oxygen
           and carbon layers as shown in \S \ref{sec:spiteplateau} below.

    \subsubsection{The  C-spallated $^7$Li  plateau}
    \label{sec:c-plateau}
    
\begin{figure*}
 \centering
\begin{tabular}{c}
\includegraphics[height=0.6\textwidth,width=0.7\textwidth,]{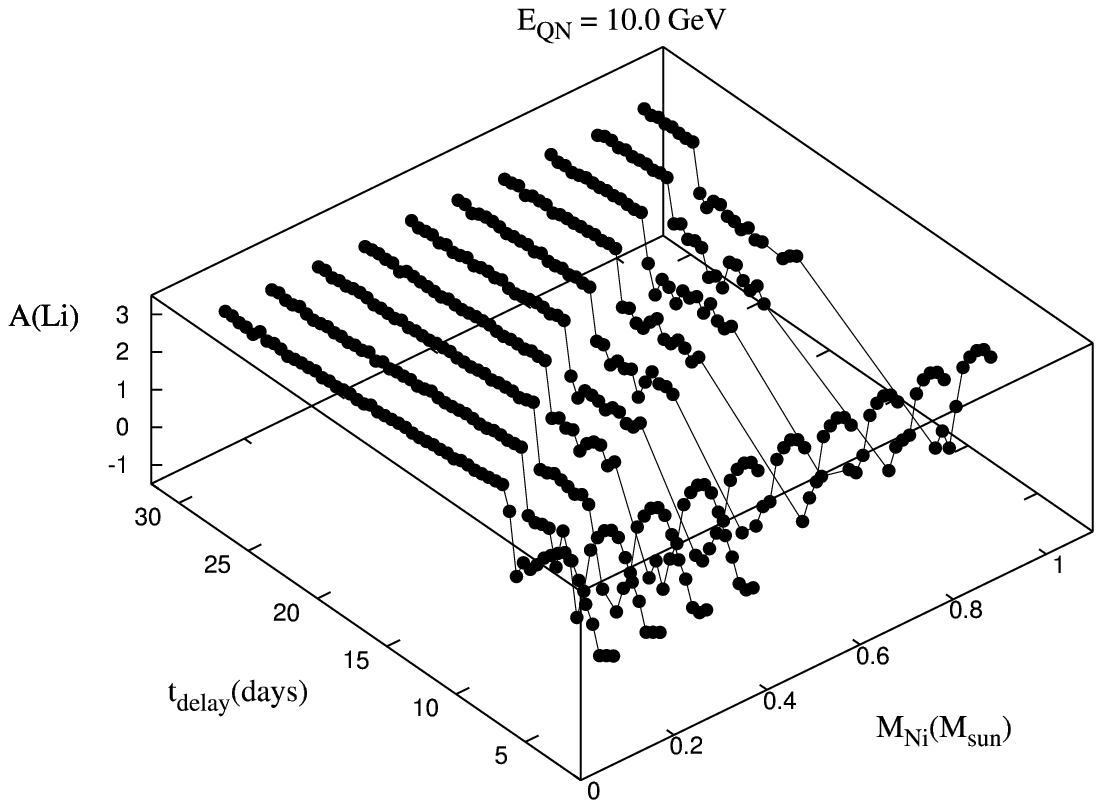}\\
\includegraphics[height=0.6\textwidth,width=0.7\textwidth,]{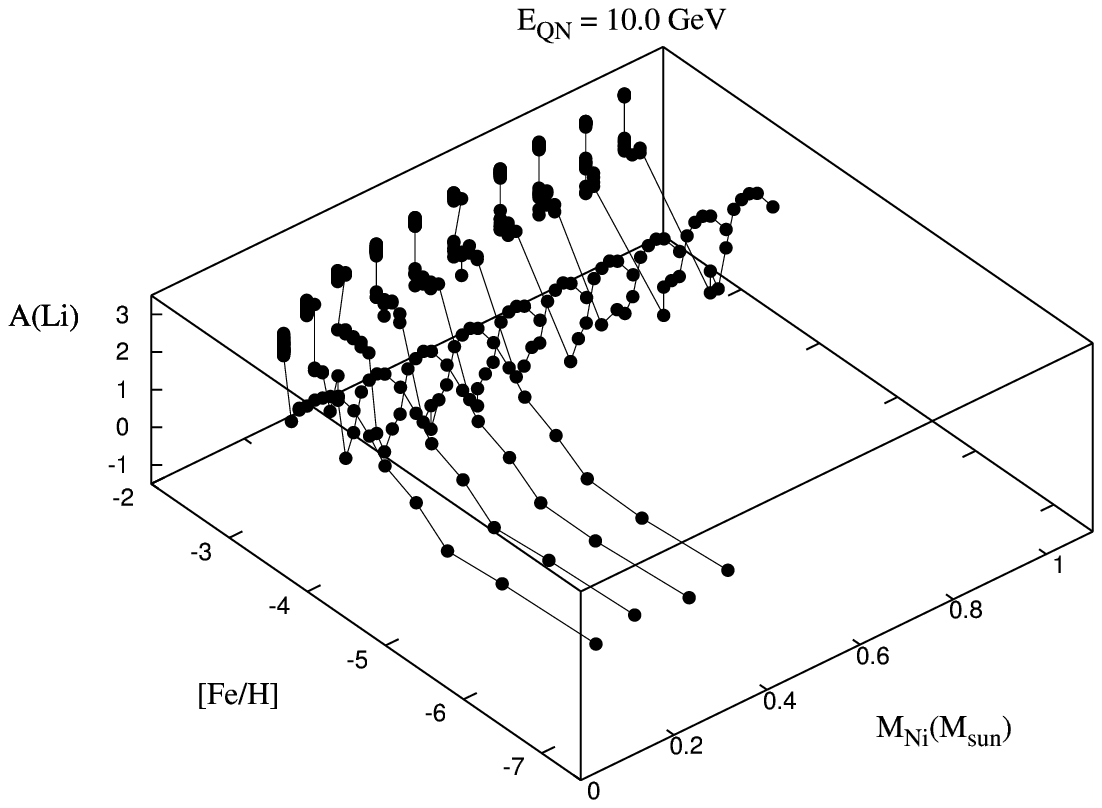}
\end{tabular}
\caption{
Same as in figure \ref{fig:3D-Li} but for $M_{\rm C, SN}=M_{\rm O, SN}= 1.5M_{\odot}$.
}
 \label{fig:super-Li}
\end{figure*}

   When $\Delta R_{\rm C} < \lambda_{\rm sp., C}$, even if enough neutrons
        with energy above $E_{\rm sp., C}\sim 0.42$ GeV make it past the O-layer, they 
     will not interact with the overlaying carbon and there will be no spallated $^7$Li and the
     corresponding plateau would   disappear.  
     From eq. (\ref{eq:Nsp}), the $N_{\rm mfp, C}>1$ condition (i.e. $\Delta R_{\rm C} > \lambda_{\rm sp., C}$) yields 
    $t_{\rm delay} < \sim 13\ {\rm days}\times  (M_{\rm C, SN}/0.15M_{\odot})^{1/2}$ for our fiducial values. This means that spallation in the
    C-layer would occur mainly when $t_{\rm outer}\sim 8\ {\rm days} < t_{\rm delay} < 13$ days which is
    a narrow window.     On the other hand, if we increase the carbon content in the PopIII-SNII ejecta as to make 
        $\Delta R_{\rm C} > \lambda_{\rm sp., C}$
        then spallation would ensue.      
          Figure \ref{fig:super-Li} shows the two plateaus again for $E_{\rm QN}=10$ GeV but
   for much higher carbon content than our fiducial values (here $M_{\rm C, SN}=M_{\rm O, SN} =1.5M_{\odot}$). 
   As can be seen in the figure, although the 
  plateau form O-spallation is close to the Spite value,  it is clearly dwarfed by  the $^7$Li
  produced in carbon layer.  For $M_{\rm C, SN}=1.5M_{\odot}$,
          the $\Delta R_{\rm C} < \lambda_{\rm sp., C}$ regime occurs when $t_{\rm delay} > 30$ days beyond  which   the
          C-spallated $^7$Li  plateau disappears.
   We find that   in PopIII-SNII with high carbon content, and in particular those with  $E_{\rm QN} \ge 10$ GeV, the $^7$Li plateau   from C-spallation is dominant and far exceeds the Spite value.  Another possibility would consist
   of  reducing oxygen content  in the  PopIII-SNII ejecta so that the primary (i.e. QN) neutrons 
experience only a very few interactions (mfps) in the oxygen layer  before making their way towards  the C-layer.
This is the least likely scenario since it seems that oxygen is produced in much larger quantity  than carbon
in PopIII-SNII ejecta (e.g. Umeda\&Nomoto 2002).

      On average, spallated neutrons    in the $E_{\rm QN}=5$ GeV case  acquire lower energies than the $E_{\rm QN}=10$ GeV case.    Thus, there is no substantial amounts of $^7$Li  spallated in the C-layer since  fewer $>0.42$ GeV neutrons manage to make past the O-layer.  The resulting plateau is very scattered and remains below Spite values; see the  panels
      labeled ``C-Target"  in Figures   \ref{fig:targets-days-5GeV} and \ref{fig:targets-FeH-5GeV}.

 \subsection{A low-scatter, $A(Li)\sim 2.2$, plateau} 
         \label{sec:spiteplateau}

\begin{sidewaysfigure}
% \begin{figure*}
 \begin{center}
\begin{tabular}{cc}
\includegraphics[scale=1.3]{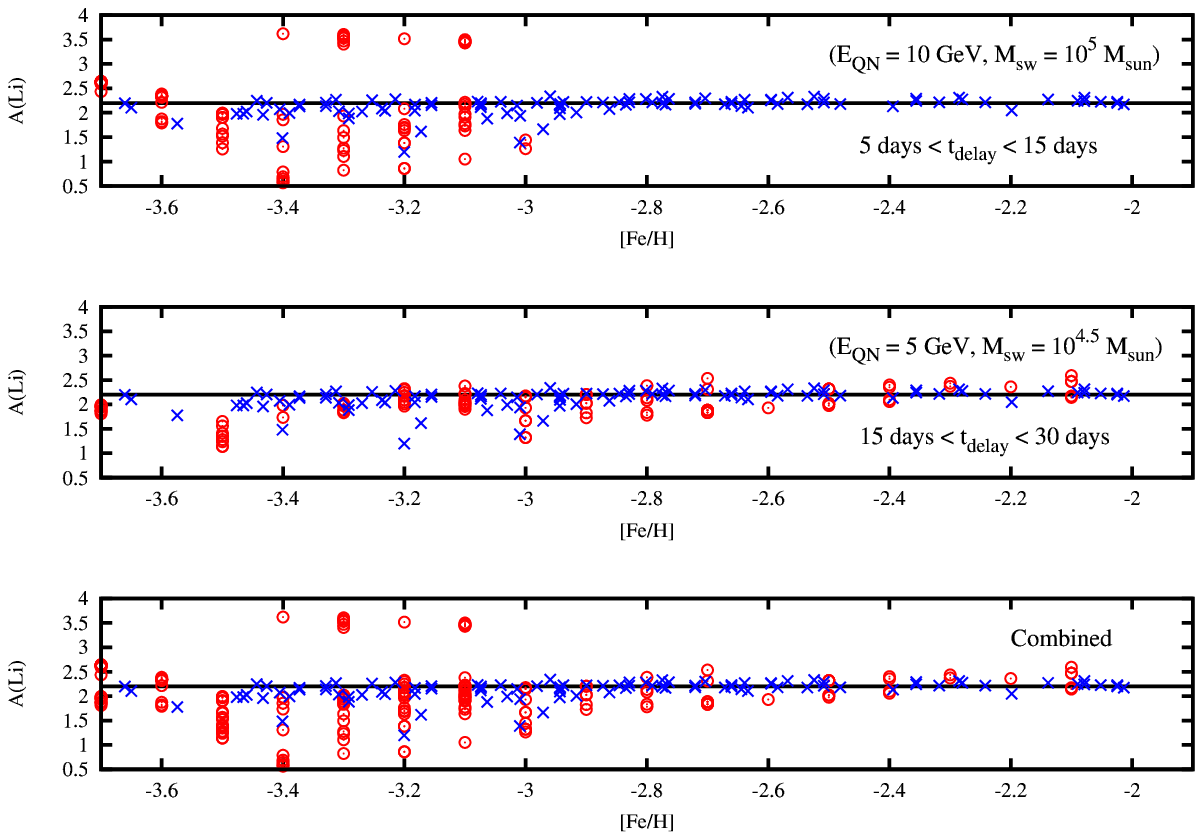} 
\end{tabular}
\caption{Top panel: $A(Li)$ versus [Fe/H]  for simulations with  $E_{\rm QN}=10$ GeV,  $M_{\rm sw}=10^5M_{\odot}$
and for  $5\ {\rm days}  <  t_{\rm delay} < 15\ {\rm days}$.   Middle panel: $A(Li)$ versus [Fe/H]  for simulations with  $E_{\rm QN}=5$ GeV, $M_{\rm sw}=10^{4.5}M_{\odot}$
and for  $15\ {\rm days}  <  t_{\rm delay} < 30\ {\rm days}$.  Bottom panel: $A(Li)$ versus [Fe/H]  resulting
from all simulations for a range in $^{56}$Ni content, $0.05 M_{\odot} \le M_{\rm Ni, SN} \le 0.5M_{\odot}$.  The crosses are measured $^7$Li abundances  in halo stars and turn-off
  stars as described in Sbordone et al. (2012;   see their figure 1 for the sources of the data).    The Horizontal line  corresponds to  $A(Li)\sim 2.2$. 
}
 \label{fig:theplateau}
  \end{center}
%\end{figure*}
 \end{sidewaysfigure}

                If we associate high $E_{\rm QN}$ dsQNe with short time delays 
               and  low $E_{\rm QN}$ dsQNe with relatively longer delays ($> 15$ days), a low-scatter plateau results as 
               shown in Figure \ref{fig:theplateau}.  The top panel shows abundances
           from simulations with $E_{\rm QN}= 10$ GeV and $M_{\rm sw} =10^{5} M_{\odot}$ with $ 5 \ {\rm days} < t_{\rm delay} < 15$ days.  The middle panel shows results from an $E_{\rm QN}=5$ GeV simulation with $M_{\rm sw} =10^{4.5} M_{\odot}$ and     for $15 \ {\rm days} < t_{\rm delay} < 30$ days.  The combined $^7$Li abundance from these
           simulations leads to a plateau with features reminiscent of the observed one.  
           To extend the plateau to higher [Fe/H] values we  had to include simulations where 
        the mass of the pristine cloud swept up by the dsQN is lower than the fiducial value of $10^{5}M_{\odot}$.
        For example,     the $M_{\rm sw} =10^{4.5} M_{\odot}$ simulations
           effectively shift the plateau to the right in Figures \ref{fig:targets-FeH-5GeV} and \ref{fig:targets-FeH-10GeV}     by  yielding [Fe/H] values  close to $\sim -2$  (see eq. \ref{eq:FeH}).      However, a decrease in $M_{\rm sw}$ increases the $A(Li)$ value (see eq. \ref{eq:ALi}) which is remedied by a decrease in $E_{\rm QN}$ which decreased the amounts of spallated $^7$Li 
           accordingly.   The high $A(Li)\sim 3.5$ values in the top panel correspond  to abundances from C-spallation. However,  the thinnest plateaus result from simulations
           which minimize  spallation in the C-layer. This  meant setting $M_{\rm C, SN} < 0.1M_{\odot}$ for our fiducial values, or  in general by taking  $M_{\rm C, SN} < M_{\rm O, SN}/10$.
           
           While  more sophisticated calculations   may be required to further constrain the parameters, 
           the QN model naturally explains this  dichotomy: 
  In general,   the reduced mass-loss in Pop. III stars  means that 
   the more massive progenitors could in principle  end up with
  the more massive cores at explosion and should thus lead to  heavier NSs.  
   Or,   if fallback is greater from the more massive QN progenitors (i.e. those closer to the  black hole limit which we 
 set here to be $40M_{\odot}$),  this should  also translate to  the formation of a  more massive compact remnant (i.e. a
  parent NS with higher core density).
  These heavier NSs  should undergo a QN explosion on shorter time delays as found by Staff et al. (2006).   The  core-collapse of the massive progenitors followed by the 
     QN explosion of the massive NS  would  likely lead to
     a more energetic SN$+$QN ejecta  which should sweep up more pristine material (i.e. $M_{\rm sw}$ should
     be higher in these dsQNe; e.g. Shigeyama\&Tsujimoto 1998). 
     Naturally, the shorter time delay (i.e. the lower [Fe/H])  dsQNe would lead to a higher scatter
      in the plateau, and its eventual meltdown for the shortest delay dsQNe, since $^7$Li will
      be produced in the inner layers from multiple targets.
       In this picture, low-mass Pop. II stars in the Spite plateau  would have originated from pristine clouds
  swept up by   dsQNe   with progenitors in the lower end of the
   mass range,  $20M_{\odot} < M_{\rm prog.} < 40 M_{\odot}$, with reduced fallback.  
As discussed later (see \S \ref{sec:discussion}),  dsQNe with short delays should be the rarest among dsQNe. The implication is that the
     resulting   Pop. II stars (those with very low  [Fe/H] and high $A(Li)$ scatter  according to our model) 
     would be the rarest among  very metal-poor stars.

\subsection{Discussion}
\label{sec:li-discussion}

        We now discuss some implications of our findings to Lithium in the Universe:

  \begin{itemize}
  
  \item  {\bf The BBN component}: In this work and in Ouyed (2013), we have neglected the
 BBN contribution in the cloud swept up by the SN$+$QN ejecta and considered only $A(Li)$ from dsQN spallation on PopIII-SNII
  ejecta.  Among plausible scenarii that could lead to the destruction of the  BBN component  in
  pristine  clouds, before they get swept up by the combined SN+QN ejecta, are: 
   
   (i) The QN provides copious amount of neutrons with some of them freely  streaming ahead of the 
     combined SN+QN ejecta.  These free QN neutrons could  destroy the BBN $^7$Li 
      (e.g., see Albornoz V\'asquez et al.  2012 and references therein for relevant nuclear reactions).
        This   might contribute to the destruction of the surrounding BBN $^7$Li before 
      material is swept up by the SN+QN ejecta -- this remains to be confirmed; 
      
   (ii)  A massive astration of $^7$Li by processing in Pop. III stars 
as suggested by Piau et al. (2006).  This model provides a natural connection to ours 
  since it  appeals to Pop. III stars in the $10 M_{\odot}$-$40M_{\odot}$ mass
 range; the favored  QN progenitor mass range.    Thus,  adopting the Piau et al. (2006) hypothesis, one could then
envision a scenario in which the BBN $^7$Li is processed and destroyed by the Pop. III star before it undergoes
a PoPIII-SNII explosion and the subsequent QN event. In this scenario,  the issue of 
overproduction of CNO elements  plaguing the Piau et al. (2006)  hypothesis 
 is  alleviated since in our scenario these would be reduced by spallation. 
 The pristine material/cloud swept up would then carry the imprint of $^7$Li 
resulting from spallation of PopIII-SNII ejecta  by dsQNe without the overproduction
of CNO elements;

 (iii) The last and the most bold proposition is the
 one where we envisage  that $^7$Li might not have  been processed during BBN and is rather a relic of primordial dsQNe spallation.

  \item  {\bf Super-Li rich candidates}:  A few super-Li rich stars have been reported in the
literature  exceeding the BBN value (e.g. the  $A(Li)= 2.87$ star in the globular cluster M4; Monaco et al. 2012).
 Other stars show abundances much above the meteoritic values of $A(Li)= 3.25$ (Deliyannis et al. 2002; Koch et al. 2012; Adam\'ow et al. 2012). Suggested  enhancement scenarios  including planetary ingestion, Hypernovae, and binary transfer (to
 cite only a few)  presumably  face challenges in accounting for such extreme levels of Li-enhancement
 (e.g. Monaco et al. 2012; Koch et al. 2012). 
 
  In our model,  such high $A(Li)$ values  result from 
    spallation    in the C layer.   C-spallation would occur for a narrow window of time delay
    ($8.3\ {\rm days} < t_{\rm delay} < 13$ days for our fiducial values) and only if enough $>0.42$ GeV
  neutrons make it past the O-layer; i.e. mainly in dsQNe with $E_{\rm QN}\ge 10$ GeV and 
   with $M_{\rm C, SN}> 0.1 M_{\rm O, SN}$
  (see discussion in \S \ref{sec:o-plateau}).    These stringent constraints  suggest that super-Li candidates (i.e.
  Pop. II low-mass stars with  $^7$Li spallated in the  C-layer in our model) should be relatively rare compared to
   normal  (i.e.  with $A(Li)\sim 2.2$ from O-spallation) candidates.

\item  {\bf Interstellar $^7$Li abundance}: Howk et al. (2012) reported measurements of interstellar $^7$Li abundance of the
Small Magellanic Cloud with values nearly equal to the BBN predictions.   We speculate that the measured value may be a signature  
 of a  gas swept up by a dsQN with important C-spallation (see discussion above).
  Interestingly the $^6$Li/$^7$Li=0.13 
 measured by Howk et al. (2012) is very close to the value predicted by Ouyed (2013; see \S 5.3 and Fig. 8 in that paper).
  The high $^6$Li/$^7$Li values are expected in our model since $^6$Li and $^7$Li would be produced with nearly
  equal proportions from spallation.
 
  \end{itemize}

 \section{Discussion and predictions}
 \label{sec:discussion}

\begin{figure*}
\centering
\includegraphics[height=0.55\textwidth, width=0.75\textwidth]{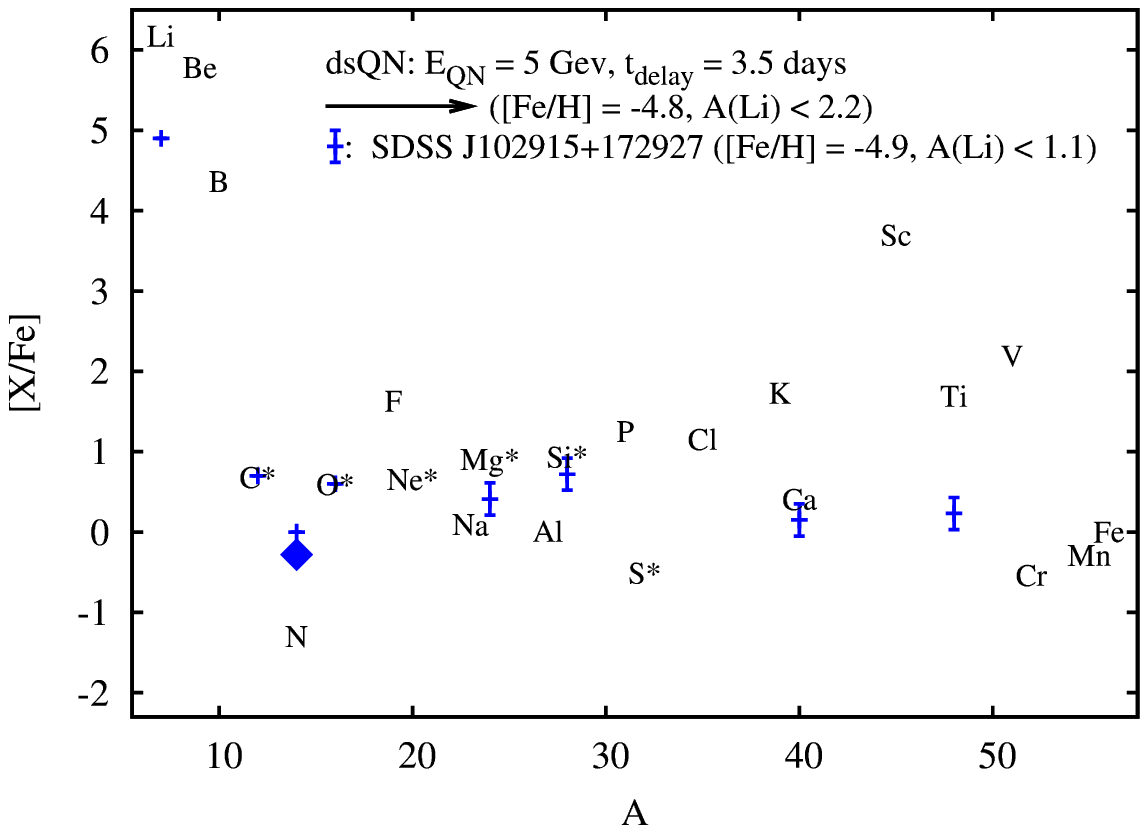} \\
\includegraphics[height=0.55\textwidth, width=0.75\textwidth]{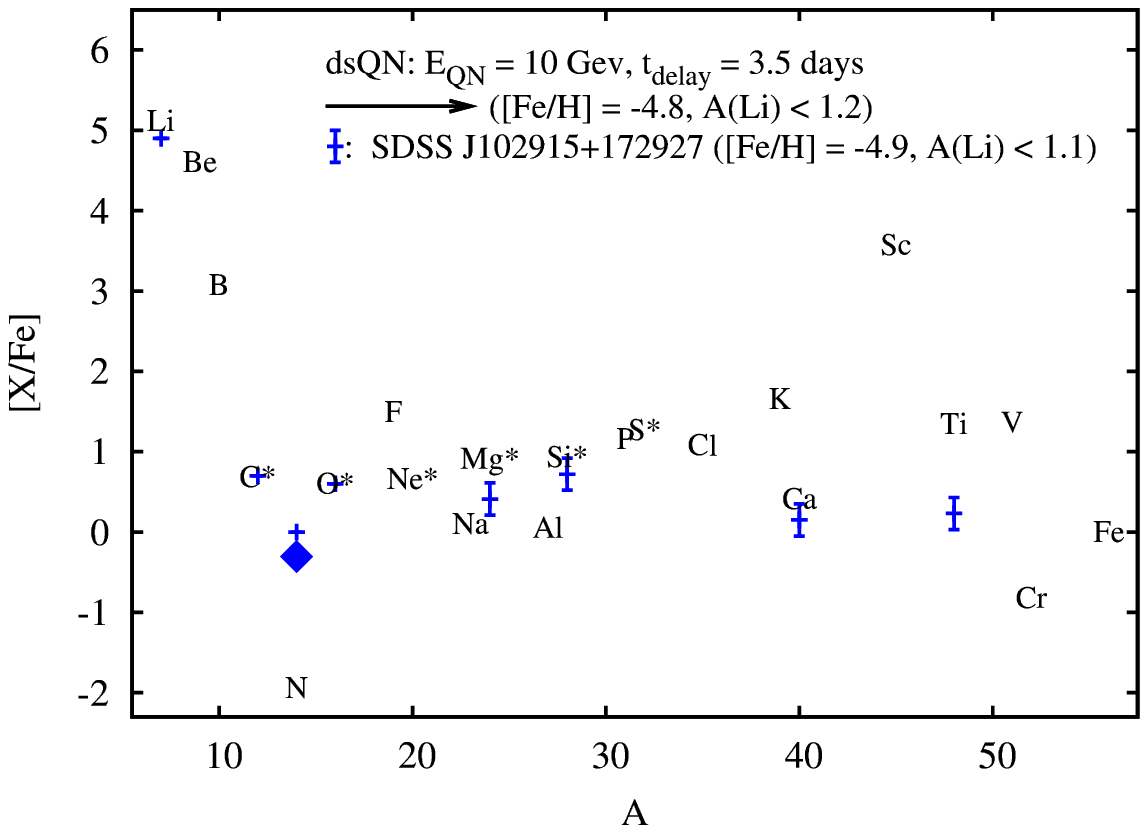} 
\caption{Relative abundances  [X/Fe] of sub-Fe spallation products (elements identified by their names) 
versus mass number $A$
from simulations with $t_{\rm delay}= 3.5$ days which  resulted in ${\rm [Fe/H]} \simeq -4.8$.  
The CN enhanced nitrogen is shown by the filled diamond. 
Top and bottom panels are for QN neutrons with $E_{\rm QN}=5$ GeV and 10 GeV, respectively. 
 SDSS J102915$+$172927's measured  abundances  are also shown, as plus ($+$) signs with the corresponding uncertainties,  
 for a comparison (plus  signs with no error bars are upper limits).  We adopted measured abundances 
  given in Caffau et al. (2011; see their Table 1) with an assumed upper limit for oxygen of $[O/Fe]\sim 0.6$.
   Best fits were found for   initial abundances in mass  of  $M_{\rm Ni, SN}= 0.1M_{\odot}$,
  $M_{\rm S, SN}=M_{\rm Si, SN}=M_{\rm Mg, SN}=M_{\rm Ne, SN}= 0.01M_{\odot}$, $M_{\rm O, SN}=0.04 M_{\odot}$ and $M_{\rm C, SN}= 0.0.02M_{\odot}$.   The mass of the cloud swept by a dsQN
 was kept at  $M_{\rm sw}=10^{5}M_{\odot}$.
}
 \label{fig:caffau}
\end{figure*}

     Below we discuss some  implications of our model followed by specific predictions:

 \begin{itemize}

\item {\bf Formation of Pop. II low-mass stars}:   
There are two  models for the transition from Pop. III to Pop. II  discussed in the literature:
 (i) Atomic fine-structure line cooling (Bromm \& Loeb 2003) where Pop. III gas clouds cool down by the emission of radiation in  de-exciting  metals. This could in principle break up the large Pop. III clouds into cores that would eventually form  Pop. II stars.  In this model, low-mass star formation can occur in gas that is enriched beyond the critical abundances of [C/H]$_{\rm crit}\sim$ -3.5 $\pm$ 0.1 dex and [O/H]$_{\rm crit}\sim$ -3 $\pm$ 0.2 dex (or  $Z>  Z_{\rm cr.}\sim 10^{-3.5}Z_{\odot}$);
  (ii) Dust-induced fragmentation (e.g. Schneider et al. 2006a). Molecular gas clouds are cooled-down by transferring their heat to the surrounding dust. This cool-down  fragments Pop. III molecular clouds into low-mass cores.  
 The dust-cooling model predicts critical metallicity $Z$ that is smaller by a factor of 10-100 than the one predicted in line cooling models.

    The discovery of   SDSS J102915$+$172927
 (Caffau et al. 2011), an ultra metal-poor (UMP) star  with ${\rm [Fe/H]} \sim -5$ (and with no excess of carbon) born from a cloud with $Z < Z_{\rm cri.}$ may be a challenge for the line cooling model and seems to rather support a fragmentation induced by dust-cooling  (see  discussion in    Silvia et al. 2010 and  Nozawa et al. 2012; see also Klessen et al. 2012).  
  
 Another possible mechanism is cloud fragmentation by the formation of high density regions (e.g. Greif et al. 2011 and  references therein).  In our model, QNe release highly dense and relativistic ejecta that can drive the surrounding cloud into compression. This cloud would be contaminated by dsQN products as well. This compression could eventually fragment the cloud into highly dense cores that turn into low-mass stars (e.g. Machida et al. 2005). An advantage of our model is that Fe-poor objects do not necessarily have low Z, since  dsQNe with low $t_{\rm delay}$ lead to spallation reactions that destroy the Ni-layer but leave outer CO-rich layers intact. Furthermore, one can have the opposite situation where PopIII-SNII C and O are spalled but not inner-layer Ni remains intact.

    Figure \ref{fig:caffau} compares relative abundances of stable isotopes
   in a pristine cloud swept up by a $t_{\rm delay}=3.5$ days dsQN to measured values
    in SDSS J102915$+$172927.  Both the $E_{\rm QN}=5$ GeV and  
    the $E_{\rm QN}=10$ GeV simulations give  very encouraging fits to measured abundances although our model
    overproduces $^{48}$Ti. As noted earlier, large abundances of Fluorine and Scandium are expected
     in our model since these rare nuclides are a natural outcome of spallation processes (see \S 6.2 in Ouyed 2013).   The $^7$Li  plotted is an upper limit. since at time delays of $t_{\rm delay} < \sim 10$ days (i.e. for [Fe/H]$< -3$), the the PopIII-SNII envelope temperature exceeds $2.5\times 10^6$ K which reduces the final abundance of spalled $^7$Li.

\item {\bf The primordial, Pop. III,  IMF}:    Our model requires no modification to
the primordial IMF (the very massive 35$M_{\odot}$-200 M$_{\odot}$  Pop. III progenitors found in simulations; e.g. Wise\&Abel 2007) and does not contradict observations that hint at contamination of  Pop. II star-forming clouds by SNII metals, where these SNII progenitors were  10M$_{\odot}$-50M$_{\odot}$  stars (e.g.  Umeda\&Nomoto 2005; Tumlinson et al. 2004). Without dsQNe, these observations contradict the primordial IMF, for the observations point at 10M$_{\odot}$-50M$_{\odot}$  progenitors, and the primordial IMF postulates much more massive progenitors. However, we have argued in previous work that the dsQN can solve this apparent contradiction (see \S 6.3 in Ouyed 2013 for more discussion and relevant references).

 \item {\bf The statistics}:   In a heavy Pop. III IMF (e.g.  see discussion in Larson 1998; see also Schneider et al. 2006b
    and references there in), dsQNe progenitors would contribute a small percentage
    of the population.   Thus, according to our model,  this scarcity of the resulting low-mass Pop. II stars in the galactic halo may be a reflection of the 
      IMF of Pop. III stars (specifically the lower mass end of the distribution). 
        Interestingly,  the  statistics  of the  known   old, [Fe/H] $< -2$ stars   are rare in the Galaxy and   represent
only a small  percentage  of the total stellar mass (see   Table 3 in Beers\&Christlieb 2005).  

	  In the QN model, a more massive progenitor NS leads to shorter time delays between the SN explosion
	   and the QN explosion.  The mass of a NS is  likely to be proportional to the mass of its progenitor star, i.e. a massive core in a SNII explosion will experience a larger fall-back than a lighter core. The more massive the fall-back is, the faster will the NS acquire a critical density that would lead into quark de-confinement and thus the QN. Therefore, massive NSs will lead to short time delays.  Because time delay is inversely proportional to the mass of the progenitor star, short $t_{\rm delay}$ dsQNe should be the rarest in a heavy Pop. III IMF, if the IMF peaks at $\sim 40M_{\odot}$.  
           In other words,  the lower the [Fe/H] (i.e. the shorter the $t_{\rm delay}$) the
        smaller the statistics of the ensuing Pop. II stars.        In this picture, Hyper-Metal Poor (HMP)
        stars     (with [Fe/H]$<-5$ in the terminology of Beers\&Christlieb 2005)    would have originated
        in pristine clouds swept up by dsQNe from progenitors on the high end in the $20M_{\odot} < M_{\rm prog.} < 40M_{\odot}$ mass range. However, this conclusion relies heavily on the exact shape of the Pop. III IMF which remains to be
         confirmed.

\item {\bf Origin of  HMP stars}:   HMP stars show very peculiar abundance enhancements of C, N, and O. 
 Astronomers have argued that these abundances may have been  a consequence of binary mass transfers (Suda et al. 2004). This hypothesis awaits the discovery of radial velocity variations in HMP stars. Furthermore, this hypothesis implies that the stellar IMF at $Z<10^{-5}Z_{\odot}$ was biased  towards intermediate mass stars (Tumlinson 2007). 
We argue that the enhanced abundance of C in HMP stars is not necessarily a consequence of binarity, but may be a signature of a dsQN. A dsQN with a sufficiently low $t_{\rm delay}$ would deplete its $^{56}$Ni layer but leave its C layer mostly intact. This would create a large ratio between C and Fe, which leads to an observed enhancement of C. In the binary accretion scenario, a large abundance of C  should be accompanied by high levels of s-process which is not always observed  (e.g. Cohen et al. 2006).
   In the dsQN model, slow neutrons capture processes and the resulting  s-process products become important 
    in dsQNe with $t_{\rm delay}$ leading to HMP stars (see \S 4 in Ouyed 2013 for more details; in particular eq.  11 in that paper). 
   
We focus on two known HMP stars, HE0107-5240 (${\rm [Fe/H]} =-5.54$; Christlieb et al. 2002) and HE1327-2326 (${\rm [Fe/H]} =-5.76$; Frebel et al. 2008).  These two stars have extremely
high enhancements of the light elements C, N, and O relative to Fe.
  Figure \ref{fig:HMPs} show a comparison of the peculiar chemical composition of HE 0107$-$5240 and
HE 1327$-$2326 to abundances from dsQNe with  for $t_{\rm delay}=2.8$ days
 and $t_{\rm delay}=2.6$ days, respectively.  The measured  relative abundances of elements, [X/Fe],  in these sources  we take from Norris et al. (2013;  see their Table 4).  
  In our model, HMPs would be produced in a situation where QN neutrons are not energetic enough to traverse the inner Ni layer and spall C and O nuclei in the outer layers (see \S 3 in Ouyed 2013). This shielding of the outer CO layers combined
  with the substantial destruction of the $^{56}$Ni layer explains the large [C/Fe] and [O/Fe] in these objects (see Figures \ref{fig:X-over-Fe}, \ref{fig:alpha-elements-5GeV} and \ref{fig:alpha-elements-10GeV}). We find a good fit to measured abundances except for $^{48}$Ti which is overproduced in our model.

 	There is a strong apparent depletion of lithium, (A(Li)$ < 1.5$) in these two HMP stars.
It's unlikely that HE1327-2326 depleted its own lithium through conventional means. A traditional mechanism for lowering the abundance of lithium would be photospheric depletion. In this picture, lithium is diffused under the photosphere and destroyed by high temperature protons. However, photospheric depletion for  HE1327-2326 is improbable, given the star's light mass and low temperature of $T_{\rm eff}\sim$ 6000 K.    It's possible that HE0107-5240 may have destroyed some $^7$Li   through photospheric depletion, but it seems unlikely that this process would lower the lithium abundance significantly below the plateau. Our model leads to the right levels of $^7$Li abundances, although the $^7$Li abundance from our simulations should be considered upper  limits since
    [Fe/H]$< -3$ for HMPs (see \S \ref{sec:o-plateau}).

       However, the extreme N abundance in 
   HE1327-2326 can be better fit   in our model  if we assume $^{14}$N to be present
   in reasonable amounts in the PopIII-SNII ejecta.  Interestingly, it has already been argued  that $^{14}$N  could
   be produced in Pop. III stars  close to the $\sim 40M_{\odot}$ limit (Heger\&Woosley 2010);  this limit
   is lower when taking rotation into account (Joggerst et al. 2010).  In our proposed picture,
    the extreme [C/Fe] and [O/Fe] values in  HE1327-2326 is already indicative of a dsQN progenitor
    with mass at the high end of the $20M_{\odot} < M_{\rm prog.} < 40M_{\odot}$ mass range. It is only fitting
    and natural to assume some $^{14}$N already present in the PopIII-SNII ejecta.  Of course,
    our model does not rule out the possibility of a companion at play in the case of HE1327-2326 given
     the initial CNO abundances we had to adopt for the fits.

  \end{itemize}

\subsection{Predictions}
  
  We now list some predictions of our model:
  
  \begin{itemize}
  
  \item  Spallation reactions in dsQNe would chemically contaminate Pop. III clouds, and therefore affect the  $^6$Li/$^7$Li ratio. A gas cloud processed by both astration and dsQNe (we assume the astration hypothesis is correct, refer to discussion in \S \ref{sec:li-discussion}) would leave different chemical signatures in daughter, Pop. II stars that a cloud only processed by astration. This  $^6$Li/$^7$Li ratio is affected because dsQN enhances $^6$Li abundance by spalling the O-layer. Therefore, the dsQNe-astration case should yield a   larger $^6$Li/$^7$Li ratio than a dsQNe-deficient case, i.e. dsQNe-astration should produce $^6$Li/$^7$Li $> 10^{-4}$. Thus, we predict that two types of low-mass, Pop. II stars should be observed: a high $^6$Li/$^7$Li  ratio star that was born from  clouds that were processed by both dsQNe and astration, and a low  $^6$Li/$^7$Li$ \sim 10^{-4}$ star produced by clouds that were only processed by astration.

    \item  We found that  dsQNe with $8.3 \ {\rm days} < t_{\rm delay} < 13$ days and $E_{\rm QN}\ge 10$ GeV, 
   lead to  spallation in the C-layer, which  resulted in $^7$Li abundances exceeding meteoritic values ($A(Li)\sim 3.25$).
    In these dsQNe, the destruction of C at the expense of lighter elements, means that super-Li
     rich low-mass stars should also show  relative depletion in carbon. However,
      $^{12}$C is also  a by-product of spallation in the $^{16}$O layer which suggest that not all 
      super-Li candidates should appear carbon poor.

  \item      QNe with $t_{\rm delay}< 2$ days  would lead to excessive (almost complete) destruction of
    $^{56}$Ni. Thus, this QNe would lead to  low-mass Pop. II stars with much lower  [Fe/H]  values than already observed HMP stars. These iron-poor stars should eventually be discovered  if QNe with extremely short time delays occur in nature. However, these dsQNe are statistically extremely rare since they would originate from 
    stars  lying the closest to the black hole limit ($\sim 40M_{\odot}$): i.e. dsQNe with  massive enough NS to trigger quark de-confinement within a small $ t_{\rm delay}$.

 \item  Pristine clouds swept up by dsQNe could produce low-mass Pop. II stars that have excessive CNO enhancements. Clouds processed by short-delayed  ($t_{\rm delay} < 2.5$ days or [Fe/H]$<-6$) dsQNe could produce these stars in binaries. In the dsQN scenario, original C and O  (and N for the more massive dsQN progenitors)  from SNII ejecta would be enhanced by spallation products. Furthermore, C, N, and O could be further enhanced by material accreted from the companion star. These CNO-enhanced objects should lead to [C/Fe], [O/Fe] and [N/Fe] values that are much greater than those measured so far in carbon-enhanced, extremely metal poor (CEMP) stars.   The discovery of these CNO-enhanced stars, in particular if they show large $^6$Li/$^7$Li ratios and large abundances of    specific spallation products  (e.g. Florine and Scandium, see  Figures \ref{fig:X-over-Fe}, \ref{fig:caffau} and \ref{fig:HMPs}), would  strongly support our model  - i.e. that the dsQN exists and it is crucial for understanding metal-poor Pop. II stars and the Spite plateau.

    \item We reiterate a previous point that the QN ejecta is not only neutron-rich,  
but rich  in $A>130$ elements as well (Jaikumar et al. 2007; see discussion in \S 6.4 in Ouyed 2013).
    A typical QN ejects on average $\sim 10^{-4}$-$10^{-3}$M$_{\odot}$
    of ejecta rich in $A>130$ elements. This ejecta's mass  is 10-100 times larger than the ejected mass of heavy element in typical core-collapse SN ejecta. A  SNII merely produces $\sim10^{-6}$M$_{\odot}$
    in heavy elements (e.g. Woosley et al. 1994 and references therein). 
    Since we estimate that on average  QNe occur at a rate of 1
     for every 100  core-collapse SNe (Leahy\&Ouyed 2009; Ouyed et al. 2009a\&b),  contamination by QNe should be equally  important if not more
    dominant.  These elements should be detectable in environments around low-mass Pop. II stars if these enviroments were contaminated by dsQNe.  A possible site for the QN $A>130$ tracers might be the ISM environment studied
     by Howk et al. (2012) in the Small Magellanic Cloud. In this environment, the super-Li $A(Li)$ value and the high $^6$Li/$^7$Li measured, might be indicative of a dsQN with C-spallation imprint.

      \item    In this work and in Ouyed (2013), we argued that extremely metal-poor 
  Pop. II low-mass stars  could have formed in pristine clouds
  swept up by primordial dsQNe.   Furthermore,  in  Ouyed et al. 2009a we showed that 
  primordial dsQNe provide enough ionizing photons to re-ionize the universe, and thus provide the $\sim 0.1$ detected optical depth in the CMB by WMAP.  Thus according to our model, 
   the end of the  epoch of re-ionization should also coincide with the epoch of formation of Pop. II low-mass stars. 
    A notion which  could be verified observationally.

  \end{itemize}

 \section{Conclusion}

        In this paper we made a case for QNe (specifically dsQNe)
        as  sources of $^7$Li in the pre-Galactic era. 
        Lithium is produced by spallation on the PopIII-SNII ejecta material.  For dsQNe with $t_{\rm delay}>  t_{\rm outer} \sim 8.3$ days, spallation in the Oxygen layer produces a $^7$Li  plateau of  $A(Li)$$\sim 2.2$ , where this lithium abundance matches with observations of Spite\&Spite (1982).     We argue for a synergy between our model and the astration model
 of Piau et al. (2006). In the astration-dsQN scenario,  $^{7}$Li that was produced in the big bang is first destroyed inside  Population III stars of a 20-40M$_{\odot}$ range.  Some of these stars then turn into dsQNe and produce Spite abundances for lithium following spallation in the O-layer.      In our model, instabilities
  in the pristine cloud swept-up by the SN+QN (i.e. dsQN) ejecta should lead to the formation
  of metal poor, low-mass stars. These stars have chemical abundances that compare well with the enhancements observed in old metal-poor dwarves. Moroever, abundances  are compatible to the values observed in turn-off stars located in globular cluster. Furthermore, our model predicts that the most iron-poor stars likely formed from  gas clouds  processed by low $t_{\rm delay}$ dsQNe. In this dsQNe picture, iron deficiency is not associated with age. In contrast to our picture, convention dictates that iron-content is proportional with cosmic age. This disentanglement between age and iron abundance calls for a paradigm shift in our understanding of iron-content in stars.

       In today's universe, QNe should be associated with massive star (20-40$M_{\odot}$)  regions.
  At low redshift,  dsQNe would  manifest themselves as superluminous SNe (SLSNe)(Ouyed et al. 2002; Leahy\&Ouyed 2008; Ouyed et al. 2010) with a distinct "double-hump" light-curve -- the first hump correspond to the core-collapse SN explosion proper and the second hump to the re-energization of the SN shell by the QN (see Ouyed et al. 2009b; Ouyed\&Leahy 2013).  Ouyed et al. (2011).  have suggested dsQNe signatures in the context of Ni-poor, Ti-rich SNe such as Cas A, and in the context of other peculiar SNe. Interestingly, the $t_{\rm delay}$ range used in the context of this paper  is close  to the $t_{\rm delay}$ range found in other dsQNe studies. The  $t_{\rm delay}$ ranges explored in this paper compare favourably to fits used for SLSNe light-curves ($ 2\ {\rm days} < t_{\rm delay} < 20$ days) (e.g. Ouyed et al. 2012; Kostka et al. 2012), and  dsQNe fits applied to chemically peculiar SNe (Ouyed et al. 2011). This similarity of $t_{\rm delay}$ ranges across QN studies may be a hint to a universal nature in the physics  underlying the QN (or a self-consistency check). These ranges in  $t_{\rm delay}$, combined with the predicted QN's gravitational wave signatures (Staff et al. 2012), should lead to dsQN signatures in future gravitational wave detectors.

    The existence of de-confined quark matter in the superdense interior of NSs is a key question that has drawn considerable attention over the past few decades. Our model postulates that this de-confinement not only exists, but triggers the most energetic explosions in the universe - the QNe. In this paper,  we have shown that the Spite plateau is created by QNe, i.e. this plateau indicates that quark matter exists inside massive NSs. Our work on the Spite plateau points at interesting relationship between the QN and the early universe: e.g. not only have we explored the QN as a mechanism for the formulation of very low-metal Pop. II stars   (Ouyed
      2013 and this work) but other studies have explored the QNe as key to the universe's re-ionization  (Ouyed et al. 2009a) as well. Other cosmological implications for the QN  have been discussed as well (Ouyed\&Staff (2013)).

\acknowledgments

I thank Jan Staff and Brian Niebergal for comments on the manuscript. I also thank Amir Ouyed for editing the manuscript.
The research of R.O. is supported by an operating
grant from the National Science and Engineering Research Council of Canada (NSERC).

%%%%%%%%%%%%% TABLESS %%%%%%%%%%%%%%%%
 \begin{table*}
\begin{center}
\caption{Average multiplicity ($\zeta$) and energy of spallated  neutrons 
for dsQNe with $t_{\rm delay}= t_{\rm outer}\sim $ 8.3 days; this is the regime
where the QN neutrons  interact with the innermost layers ($^{56}$Ni, $^{32}$S, $^{28}$Si).  
   For  this time delay, $N_{\rm mfp, A_T}\sim 1$  which means that the thickness of each target layer  is   equivalent to roughly   one spallation mfp (see eq. \ref{eq:Nsp}). 
 Spallation  stops at the $^{28}$Si layer since the subsequent generation of neutrons exit the $^{32}$S layer with energies  $\sim 0.15$ GeV
 which is below critical for spallation to ensue in the overlaying layers $^{24}$Mg layer ($E_{\rm sp., Mg}\sim 0.18$ GeV; see discussion following eq. \ref{eq:zeta}).}
 
\begin{tabular}{|c||c|c|c||c|c|c|}\hline

 &  \multicolumn{3}{|c|}{Primaries ($E_0=E_{\rm QN}=10$ GeV)}   &  \multicolumn{3}{|c|}{Primaries ($E_0=E_{\rm QN}=5$ GeV)}\\
 
 \hline
%\cmidrule(lr){2-4}  \cmidrule(lr){5-7}\hline

   $A_{\rm T}$  & $^{56}$Ni & $^{32}$S &  $^{28}$Si   &     $^{56}$Ni &  $^{32}$S &  $^{28}$Si   \\\hline
   $\zeta_{\rm av., i}$  &     $\sim 13.1$ & $\sim 3.6$  & $\sim 1.4$  &        $\sim 11.3$ & $\sim 2.8$  & $\sim 1.1$   \\
   
   $E_{\rm av., i} = \frac{E_{{\rm av.,  i-1}}}{\zeta_{av., i}}$\  (GeV)  &     $\sim 0.77$ & $\sim 0.21$  & $\sim 0.15$  &        $\sim 0.44$ & $\sim 0.16$  & $\sim 0.15$   \\
   
    $^{\dagger}A_{\rm peak, i}\simeq A_{\rm T}-\zeta_{av., i}$  &     $\sim ^{44}$Ti & $\sim ^{28}$Si  & $\sim ^{27}$Al  &        $\sim ^{45}$Sc & $\sim ^{28}$Si  & $\sim ^{27}$Al   \\    
    
     $^{\ddagger}\zeta_{\rm net, i} = \zeta_{\rm av., i-1}\times  \zeta_{\rm av., i}$  &     $\sim 13.1$ & $\sim 47.2$  & $\sim 66$  &        $\sim 11.3$ & $\sim 31.7$  & $\sim 34.8$

                                                                    \\\hline
\end{tabular}\\
$^{\dagger}$The resulting spallation products would acquire a normal distribution peaking at  atomic weight   $A_{\rm peak, i}\sim A_{\rm T}-\zeta_i$.\\
$^{\ddagger}$  The net multiplicity ($\zeta_{\rm net}$; see eq. 4 in Ouyed 2013) translates to a total spallation neutrons and protons which amounts to $M_{\rm n+p}= \zeta_{\rm net}M_{\rm QN}\sim$ 0.05-0.1 M$_{\odot}$.\\
 
 \label{tab:ni-spallation}~\\
\end{center}

\end{table*}
 
\clearpage  

\begin{sidewaystable}[ht]
% \begin{table*}
\begin{center}
\caption{Average multiplicity ($\zeta$) and energy of spallated  neutrons in the oxygen ($^{16}$O) layer 
for dsQNe with $t_{\rm delay}> t_{\rm outer}$ (here $t_{\rm delay}=12$ days); this is regime
where the QN neutrons barely interact with the inner layers and proceed directly to the oxygen layer. 
   The thickness of the oxygen layer  (with $M_{\rm O, SN}=1.5M_{\odot}$) is equivalent to  $N_{\rm sp., O}\sim 10$   sub-layers each of radial thickness 
 or the order of a spallation mfp (see eq. \ref{eq:Nsp}). 
 Spallation  in the oxygen layer ceases when the neutron energy is below the threshold value for O-spallation
($E_{\rm sp., O}\simeq 0.27$ GeV) or when the neutrons run out of target material (i.e.
after  $N_{\rm sp., O}\sim 10$ mfps).}

\begin{tabular}{|c||c|c|c|c|c||c|c|c|c|c|}\hline

 &  \multicolumn{5}{|c|}{Primaries ($E_0=E_{\rm QN}=10$ GeV)}   &  \multicolumn{5}{|c|}{Primaries ($E_0=E_{\rm QN}=5$ GeV)}\\
 
 \hline
%\cmidrule(lr){2-6}  \cmidrule(lr){7-11}

 $A_{\rm T} = 16$   & mfp 1 & mfp 2 &  mfp 3 &  mfp 4 &  mfp 5  &    mfp 1 &  mfp 2 &  mfp 3 &  mfp 4 &  mfp 5 \\\hline
   $\zeta_{\rm av.,  i}$  &     $\sim 3.8$ & $\sim 2.7$  & $\sim 2$  & $\sim 1.5$  & $\sim 1.1$ 
                               &     $\sim 3.2$ & $\sim 2.4$  & $\sim 1.7$  & $\sim 1.3$  & $\sim 1.1$ \\
   
   $E_{\rm  av.,  i} = \frac{E_{{\rm av., i-1}}}{\zeta_{av., i}}$\  (GeV) & $\sim$ 2.7 & $\sim$1.0 & $\sim$0.5 & $\sim$0.34 & $\sim$0.28 
                                                                                         & $\sim 1.6$ & $\sim 0.7$  & $\sim 0.4$  & $\sim 0.3$  & $\sim 0.28$    \\

  $^{\dagger}A_{\rm peak, i}\simeq 16-\zeta_{av., i}$ &  $^{12}$C  &   $^{13}$C  &   $^{14}$N  &    $^{14}$N  &  $^{15}$O    
                                                                  & $^{13}$C  &   $^{14}$N  &   $^{14}$N  &    $^{15}$O  &  $^{15}$O \\
                                                                  
                                                                  $^{\ddagger}\zeta_{\rm net, i} = \zeta_{\rm i-1}\times  \zeta_i$ & $\sim 3.8$  &   $\sim 10.3$  &   $\sim 20.6$  &    $\sim 31$  & $\sim  34$     
                                                                                           & $\sim 3.2$  &   $\sim 7.7$  &   $\sim 13$  &    $\sim 17$  & $\sim  19$                                                   
                                                                    \\\hline
\end{tabular}\\
$^{\dagger}$The resulting spallation products would acquire a normal distribution peaking at  atomic weight   $A_{\rm peak, i}\sim 16-\zeta_i$.\\
$^{\ddagger}$  The net multiplicity ($\zeta_{\rm net}$; see eq. 4 in Ouyed 2013) translates to a total spallation neutrons and protons which amounts to $M_{\rm n+p}= \zeta_{\rm net}M_{\rm QN}\sim$ 0.02- 0.04 M$_{\odot}$.
 \label{tab:o-spallation}
\end{center}
%\end{table*}
\end{sidewaystable}
%%%%%%%%%%%%% TABLES %%%%%%%%%%%%%%%%

%%%%%%%%%%%%% FIGURES %%%%%%%%%%%%%%%%

% \begin{landscape}
\clearpage
 
 \clearpage
%\end{landscape}

 \clearpage

 \clearpage

  \clearpage

  \clearpage

 \clearpage
 
 \clearpage

 \clearpage

 \clearpage

 \clearpage
\begin{figure*}
\centering
\includegraphics[height=0.55\textwidth, width=0.75\textwidth]{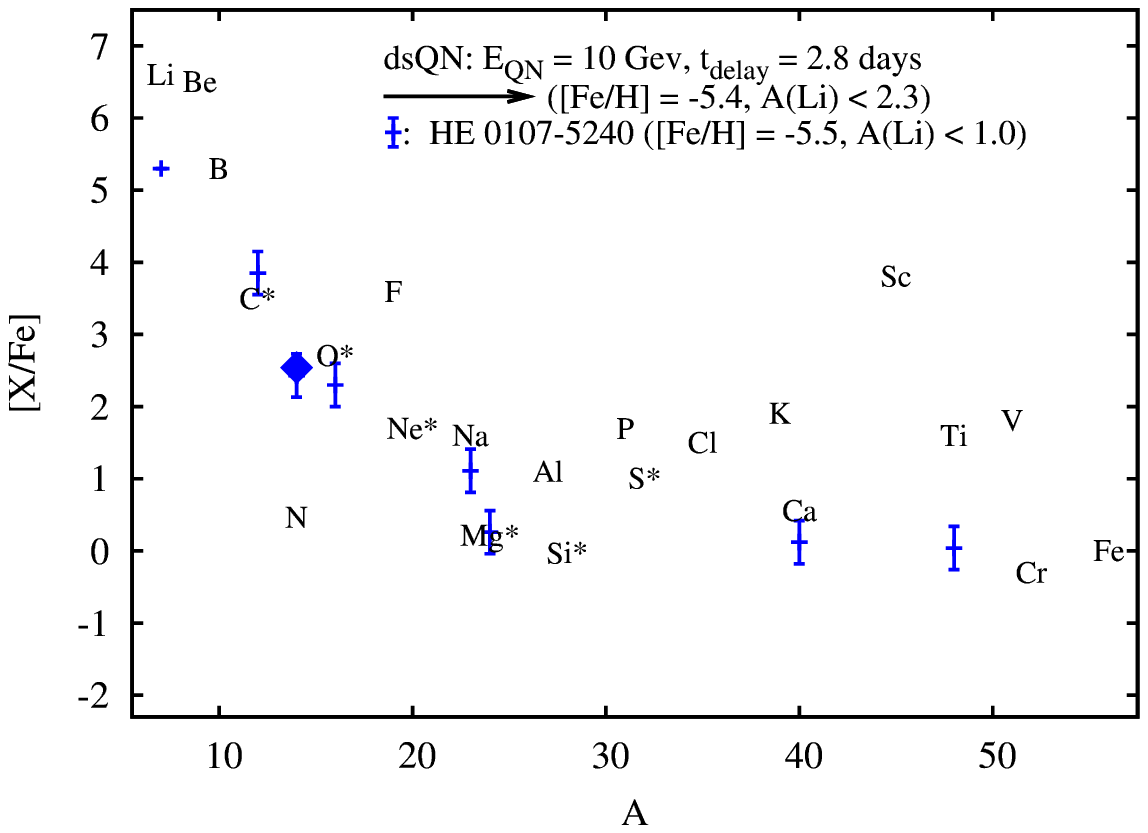} \\
\includegraphics[height=0.55\textwidth, width=0.75\textwidth]{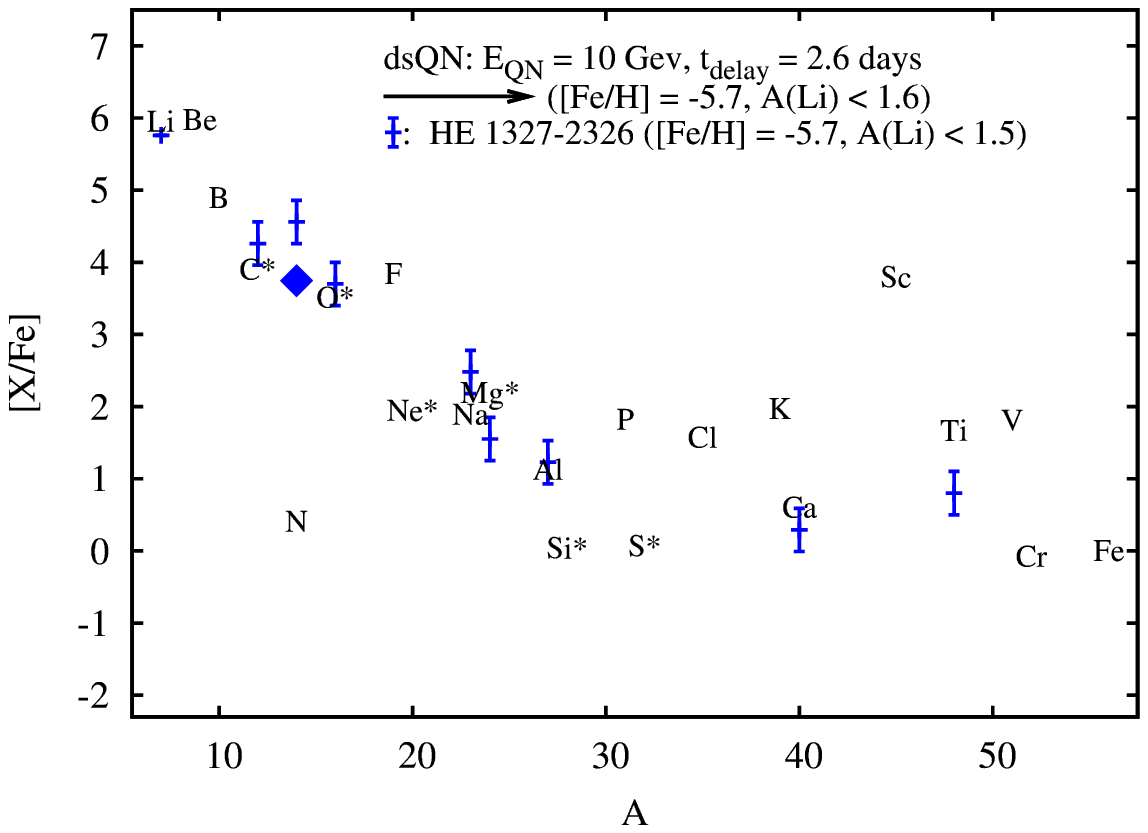} 
\caption{{\bf Top panel}: Relative abundances  [X/Fe] of sub-Fe spallation products (elements identified by their names) 
versus mass number $A$
from simulations with $t_{\rm delay}= 2.8$ days and $E_{\rm QN}=10$ GeV  which  resulted in ${\rm [Fe/H]} \simeq -5.4$ and $A(Li) < 2.3$.  The CN enhanced nitrogen is shown by the filled diamond. 
 HMP-HE-0107-5240's measured  abundances  are also shown, as plus ($+$) signs with the corresponding uncertainties,  
 for a comparison (plus signs with no error bars are upper limits).   The fit was found for   initial abundances in mass  of  $M_{\rm Ni, SN}= M_{\rm S, SN}= M_{\rm Si, SN} =0.01M_{\odot}$,  $M_{\rm Mg, SN}=M_{\rm Ne, SN}= 0.02M_{\odot}$, $M_{\rm O, SN}= 1.0 M_{\odot}$ and $M_{\rm C, SN}= 3.0 M_{\odot}$. 
   The CN enhanced nitrogen was obtained  for a 1\% enhancement factor;   {\bf Bottom  panel}:  Relative abundances  [X/Fe] of sub-Fe spallation products (elements identified by their names) 
versus mass number $A$
from simulations with $t_{\rm delay}= 2.6$ days and $E_{\rm QN}=10$ GeV  which  resulted in ${\rm [Fe/H]} \simeq -5.7$ and $A(Li) < 1.5$.  
 HMP-HE-0107-5240's measured  abundances  are also shown, as plus ($+$) signs with the corresponding uncertainties,  
 for a comparison.   The fit was found for   initial abundances in mass  of  $M_{\rm Ni, SN}= M_{\rm S, SN}= M_{\rm Si, SN} =0.01M_{\odot}$,  $M_{\rm Mg, SN}=M_{\rm Ne, SN}= 0.02M_{\odot}$, $M_{\rm O, SN}= 3.5 M_{\odot}$ and $M_{\rm C, SN}= 3.5 M_{\odot}$. 
   Both cases were run for  an $M_{\rm sw}=10^{5}M_{\odot}$  cloud swept by the dsQN.  The measured   abundances  were taken from Norris  et al. (2013). 
}
 \label{fig:HMPs}
\end{figure*}

%%%%%%%%%%%%% FIGURES %%%%%%%%%%%%%%%%

\end{document}